\documentclass[aps,prl,twocolumn,amsmath,amsfonts,notitlepage]{revtex4-1}
\usepackage{graphicx}
\usepackage{xcolor}

\DeclareGraphicsExtensions{.eps,.pdf,.PDF}

\DeclareMathOperator\erfc{erfc}

\begin{document}

\title{%Complete random matrix classification of SYK models with odd number of Majoranas
Periodic Table of SYK and supersymmetric SYK }
%and Spectrum Hard Edges}

\author{Fadi Sun}
\affiliation{
Department of Physics and Astronomy,
Mississippi State University, MS 39762, USA}
\affiliation{
Key Laboratory of Terahertz Optoelectronics, Ministry of Education
and Beijing Advanced innovation Center for Imaging Technology,
Department of Physics, Capital Normal University, Beijing 100048, China}
\affiliation{
Kavli Institute of Theoretical Physics,
University of California, Santa Barbara, Santa Barbara, CA 93106, USA}
\author{Jinwu Ye}
\affiliation{
Department of Physics and Astronomy,
Mississippi State University, MS 39762, USA}
\affiliation{
Key Laboratory of Terahertz Optoelectronics, Ministry of Education
and Beijing Advanced innovation Center for Imaging Technology,
Department of Physics, Capital Normal University, Beijing 100048, China}
\affiliation{
Kavli Institute of Theoretical Physics,
University of California, Santa Barbara, Santa Barbara, CA 93106, USA}

\date{\today}

\begin{abstract}
We develop a systematic and unified random matrix theory to classify
Sachdev-Ye-Kitaev (SYK) and supersymmetric (SUSY) SYK models
and also work out the structure of the energy levels in one periodic table.
The SYK with even $q$- and SUSY SYK with odd $q$-body interaction,
$N$ even or odd number of Majorana fermions
are put on the same footing in the minimal Hilbert space,
$N\pmod 8$ and  $q\pmod 4$ double Bott periodicity are identified.
Exact diagonalizations are performed to study
both the bulk energy level statistics and hard edge behaviours.
A new moment ratio of the smallest positive eigenvalue
is introduced to determine hard edge index efficiently.
Excellent agreements between the ED results
and the symmetry classifications are demonstrated.
Our complete and systematic methods can be transformed
to map out more complicated periodic tables of SYK models
with more degree of freedoms, tensor models
and symmetry protected topological phases. Possible classification of charge neutral quantum black holes are hinted.
\end{abstract}

\maketitle

%The $\mathcal{N}=1$ supersymmetric hybrid of $q=2$ and odd $q>2$ SYK

{\color{blue}\textit{1. Introduction.}}---
Extensive research efforts are currently being devoted to investigate
quantum chaos and quantum information scramblings
in Sachdev-Ye-Kitaev (SYK) model
\cite{Sachdev-Ye,Georges2001,KitaevTalks,JT-2,Sachdev2015,Ye}
and its %various
supersymmetric generalization \cite{Fu2017}.
A salient feature of the SYK models is that it is quantum chaotic in both
early time (Ehrenfest time)
\cite{Polchinski2016,Maldacena2016,Gross2017,Qi,Bagrets2016,liu1,liu2}
and late time (Heisenberg time)
\cite{Fu2016,You2017,Cotler2017,Li2017,Kanazawa2017,Garcia2016,Garcia2018,hybrid}.

In the early time, an exponential grow in the out of time correlation
functions can be used to characterize the quantum information scramblings.
It was found that the SYK models show the maximal quantum chaos
with the largest possible Lyapunov exponent $\lambda_L=2\pi/\beta$
saturating the quantum chaos bound \cite{chaosbound}.
This feature ties that of the quantum black holes
which are the fastest quantum information scramblers in Nature.
This remarkable feat suggests that the SYK models may be
a boundary theory of some sort of bulk dilaton gravity theory
such as the well known 2d Jackiw-Teitelboim gravity \cite{JT-1,JT-2}.

In the late time, despite its sparse nature in randomness, the SYK models
can still be described by random matrix theory (RMT) \cite{Fu2016,You2017,Cotler2017,Li2017,Kanazawa2017,Garcia2016,Garcia2018,hybrid,caution}.
The RMT has also been employed to study the quantum chaotic
behaviours of event horizon fluctuations of black holes \cite{Cotler2017}.
%The quantum chaos in the SYK models are due to the quenched disorders.
%It was found that the energy level statistics of the
%Majorana fermion SYK with $q=4$ can be described by the 3-fold way
%Wigner-Dyson (WD) distributions in a $N \pmod 8$ periodicity
%\cite{You2017,Cotler2017,Garcia2016}.
%Also different dip-ramp-plateau behaviours of SFF is found to be appear in the same periodicity
%\cite{Cotler2017}.
%These different statistic behaviours arise from the fact
%that SYK model with different $N$ belongs to different RMT classes.
The RMT classifications of the SYK models
have been applied to $q\pmod 4=0$ for all $N$ \cite{You2017,Cotler2017,Garcia2016,hybrid}
and $q\pmod 4=1,2,3$ for even $N$ case \cite{Li2017,Kanazawa2017} only.
It was generally believed that the  odd $ N $ case need to be treated quite differently than the even $ N $ case.
In fact, $N$ odd case was first discussed in the context of the 1d SPT phase \cite{Fidkowski2011},
and then applied to the SYK model with $q=4$ in \cite{You2017,hybrid}.
The authors developed a special procedure to treat $N$ odd case
by adding an extra Majorana fermion at infinity into the system.
While such a procedure is not necessary in the even $ N $ case.

In this work, by employing the Clifford algebra representation of Majorana fermions in a minimal Hilbert space,
we develop a systematic and unified random matrix theory to classify
SYK models with generic $q$-body interaction and general $N$-sites.
After identifying complete set of conserved (or chiral) quantities and anti-unitary (or unitary) symmetry operators,
we also work out the fine structure of the energy levels explicitly.
The SYK with even $ q $ and SUSY SYK with odd $ q $, $ N $ even or odd are treated on the same footing.
There are 9 classes, except class AIII(chGUE) with $ N \pmod 8 $ and  $q \pmod 4=0$ double Bott periodicity
identified in one periodic table  \ref{tab:SYKq}.
Our systematic approach not only reproduces the previously known results
in a much more efficient way, but also lead to new results
on $q\pmod 4=1,2,3$ with odd $ N $, thus complete the whole periodic table for the SYK models.
Therefore it provides additional deep and global insights into the family of SYK models showing maximal quantum chaos.

%    After exhausting all the symmetries operators,
%    we classify SYK models for all $N$, $q$ cases,
%    and thus complete the periodic table for $\mathcal{N}=0,1$ Majorana SYK models.
%    We complete the eight-fold periodicity in $N$ for $q\pmod 4=1,2,3$ cases,
%	and obtain a periodic table for $\mathcal{N}=0,1$ Majorana SYK models.

For the 7 RMT classes with a spectral mirror symmetry,
we derive a Wigner-like surmise for the probability function for the lowest positive energy level,
which can be used to describe the quantum chaotic behaviours near the hard edge.
A new moment ratio $A$ ( See SM-A ) is introduced to determine  the hard edge exponent effectively.
When combining with the $r$-parameter \cite{Atas2013,You2017,hybrid} used to
determine the bulk energy level statistics (ELS), the parameter pair $ ( r, A) $ can be used to determine the 10-fold way uniquely
and efficiently. Exact diagonalizations (EDs)  are performed to study both the bulk
%energy level statistics
ELS and hard edge behaviours.
Excellent agreements are found between the ED results and the symmetry classifications.
We also establish intricate and constructive  connections between the unified minimum  scheme and various other extended schemes
in both classification and  degeneracy counting in odd $ N $ case.
Our unified scheme and the new moment ratio $A$ can be easily transformed 
to study other SYK-like models with more symmetries,  tensor models and classifications of SPT phases.

%Fine tuning the table
\renewcommand{\arraystretch}{1.3}%Default value is 1
\renewcommand\tabcolsep{4pt}%Default value is 0pt

\begin{table}%[!htb]
\caption{
The periodic table of SYK for even $ q >2 $
and SUSY SYK for odd $q >1 $.
Only AIII in the 10-fold way is missing \cite{un}.
All the results in the table were achieved in a unified minimum scheme.
The results for $ q\pmod4 =0 $ agree with those in \cite{You2017,hybrid}  for even $ N $
and also odd $ N $ in the extended scheme ( see SM C ),
for  $q\pmod4=1,2,3$ for even $ N $ agree with those in \cite{Li2017,Kanazawa2017}.
Those for $q\pmod4=1,2,3$ with odd $N$ are new. In the listed degeneracy, $ 1+1 $  means the
two degenerate energy levels in the two opposite parities.
%The symmetry class is labeled by its Cartan name.
%$P$ and $R$ are relevant symmetry operators.%, see main text.
}
\begin{tabular}{ c|cccccccc }
\toprule
    $N\pmod 8$	&0	&1	&2	&3	&4	&5	&6	&7	\\ \hline
$P^2$ value	&$+$	&$+$	&$+$	&$-$	&$-$	&$-$	&$-$	&$+$	\\
$R^2$ value	&$+$	&$+$	&$-$	&$-$	&$-$	&$-$	&$+$	&$+$	\\ \hline
$q\pmod4=0$	&AI	&AI	&A	&AII	&AII	&AII	&A	&AI	\\
$Q_q$
degeneracy	&1	&1	&1+1	&2	&2	&2	&1+1	&1	\\ \hline
$q\pmod4=2$	&D	&D	&A	&C	&C	&C	&A	&D	\\
$Q_q$	
degeneracy	&1	&1	&1	&1	&1	&1	&1	&1	\\  \hline
$q\pmod4=1$	&BDI    &AI	&CI	&C	&CII	&AII	&DIII	&D	\\
$Q_q$
degeneracy	&1	&1	&1	&1	&2	&2	&2	&1	\\
$H$
degeneracy	&2	&1	&2	&2	&4	&2	&4	&2	\\ \hline
$q\pmod4=3$	&BDI	&D	&DIII	&AII	&CII	&C	&CI	&AI	\\ %\hline
$Q_q$
degeneracy	&1	&1	&2	&2	&2	&1	&1	&1	\\
$H$
degeneracy	&2	&2	&4	&2	&4	&2	&2	&1	\\
\toprule
\end{tabular}
\label{tab:SYKq}
\end{table}

{\color{blue}
\textit{2. A unified and systematic scheme to classify Majorana SYK models.}}---
Consider a generic all-to-all $q$-body Majorana interacting Hamiltonian
\begin{align}
    Q_q=i^{\lfloor q/2\rfloor}\sum_{i_1<\cdots<i_q}
	C_{i_1\cdots i_q}\chi_{i_1}\chi_{i_2}\cdots\chi_{i_q}
	%+\sum_i D_i\chi_i
\label{eq:Qq}
\end{align}
where $C_{i_1\cdots i_q}$ is real and satisfy
the Gaussian distributions with mean $\langle C_{i_1\cdots i_q}\rangle=0$
and variance $\langle C_{i_1\cdots i_q}^2\rangle=J_\text{SYK}^2(q-1)!/N^{q-1}$ for even $q$
or $\langle C_{i_1\cdots i_q}^2\rangle=J_{\mathcal{N}=1}(q-1)!/N^{q-1}$ for odd $q$.
The overall factor $i^{\lfloor q/2\rfloor}$ ensures the Hermitian of $Q_q$,
and $\lfloor q/2\rfloor$ (integer floor) denotes the biggest integer smaller than $q/2$.
When $q$ is even, $Q_q$ is a bosonic operator
which is the SYK Model with $q$-fermion interactions \cite{KitaevTalks,JT-2,Maldacena2016};
when $q$ is odd, $Q_q$ is a fermionic operator
which is the supercharge of $\mathcal{N}=1$ supersymmetric SYK model \cite{Fu2017,Li2017,Kanazawa2017}.
Below we choose $J_\text{SYK}=1$ and $J_{\mathcal{N}=1}=1$ when performing numerical calculations.
%\cite{You2017,Cotler2017},\cite{Li2017,Kanazawa2017},\cite{KitaevTalks,Fidkowski2011}

The Hamiltonian Eq.\eqref{eq:Qq} contains $N$ Majorana fermions
satisfying the Clifford algebra $\{\chi_i,\chi_j\}=2\delta_{ij}$
and $\chi_i^\dagger=\chi_i$,
thus admitting a $2^{\lfloor N/2\rfloor}$-dimensional matrix representations
\cite{Clifford}.
In this representation, one can choose $\chi_i$ with odd $i$ to be real and symmetric,
and $\chi_i$ with even $i$ to be pure imaginary and skew-symmetric \cite{footnote1}.
%[Notice: one can not choose odd $i$ to be pure imaginary and even $i$ to be real when $N$ is odd,
%becase the number of real matrices equal to the number of real matrices plus one.]
%
%
By collecting real and imaginary represented Majoranas, one can define
two anti-unitary particle-hole symmetry operators:
$P=K\prod_{i=1}^{\lceil N/2\rceil}\chi_{2i-1}$
and
$R=K\prod_{i=1}^{\lfloor N/2\rfloor}i\chi_{2i}$,
%\footnote
%{[This is genenalization of P and R in JHEP05(2017)118.]}
where $\lceil N/2\rceil$ (integer ceiling) denotes the smallest integer greater than $N/2$, therefore
$\lfloor N/2\rfloor+\lceil N/2\rceil=N$ holds for any integer $N$.

It can be shown that
$P\chi_i P^{-1}=-(-1)^{\lceil N/2\rceil}\chi_i$
and
$R\chi_i R^{-1}=(-1)^{\lfloor N/2\rfloor}\chi_i$,
thus
\begin{subequations}
    \begin{eqnarray}
    PQ_qP^{-1}&=&(-1)^{\lfloor q/2\rfloor}(-1)^{q(1+\lceil N/2\rceil)}Q_q,\\
    RQ_qR^{-1}&=&(-1)^{\lfloor q/2\rfloor}(-1)^{q\lfloor N/2\rfloor}Q_q,
    \end{eqnarray}
\label{eq:CMR}
\end{subequations}
One can also find their squared values
\begin{align}
    P^2=(-1)^{\lfloor \lceil N/2\rceil/2 \rfloor},\quad
    R^2=(-1)^{\lfloor 3\lfloor N/2\rfloor/2 \rfloor}\>.
\label{eq:squared}
\end{align}

Multiplying two equations in Eq.\eqref{eq:CMR} leads to the following unified classification:
When $q(1+N)$ is odd,
one of $P$, $R$ commutes with $Q_q$ and the other anti-commutes with $Q_q$,
thus $\Lambda=PR$ is a chirality operator satisfying $\{\Lambda,Q_q\}=0$;
when $q(1+N)$ is even, $P$ and $R$ either both commute with $Q_q$ or both anti-commute with $Q_q$,
thus $\Lambda=PR$ is a conserved quantity (fermion number parity)
satisfying $[\Lambda,Q_q]=0$
if $N$ is even, or an identity operator if $N$ is odd.

In the above case when $\Lambda$ is a conserved parity,
namely, when $ (q,N)=$ (even, even),
the Hamiltonian can be block-diagonal decomposed into even and odd parity sectors.
Using the fact $ PR=(-1)^{\lceil N/2\rceil\lfloor N/2\rfloor}RP$, % and $KP$, $KR$ are real,
one can find out the relation between the anti-unitary operators and the conserved parity:
$[P, \Lambda]=[R, \Lambda]=0 $ if $N\pmod 8=0,4 $, which means $P$ and $R$ preserve the parity;
$\{P, \Lambda \}=\{R, \Lambda \}=0$ if $N\pmod 8=2,6$, which means $P$ and $R$ swap the parity.
If $\Lambda$ is a chirality operator,
namely, when $ (q, N)=$ (odd, even),
which is also a parity operator,
then there is no need to worry about these commutation relations.
Of course, if $\Lambda=1 $,
namely, when $(q,N)=$ (odd, odd) or $(q,N)=$ (even, odd),
then it can be ignored anyway.

The classification can be systematically done by
evaluating  Eq.\eqref{eq:CMR}, Eq.\eqref{eq:squared} and their relations with $ \Lambda $ presented above.
The periodicity of the classification can be directly inferred
from that of  Eq.\eqref{eq:CMR} and Eq.\eqref{eq:squared}
which are invariant under $q\to q+4$ or $N\to N+8$.
So the Bott periodicity in $q$ and in $N$ is $ 4 $ and $8$ respectively.
So in the following, one need only apply the scheme to $q\pmod 4$ and $N\pmod 8$ cases.

{\color{blue}\textit{3. SYK models with $q\pmod 4=0$.}}---
Since $q(1+N)$ is even, $[P,Q_q]=[R,Q_q]=0$, $\Lambda=(-1)^F $ is a conserved  parity  when $N$ is even
or $ \Lambda=1 $ when $N$ is odd.
Depending on the commutation relation between $P$ and $\Lambda$, the value of $P^2$, one reach the following classification:
When $N\pmod 8=2,6$, both $ P $ and $ R $ swap the parity, thus $Q_q$ is in Class A(GUE),
the energy level degeneracy is $1+1$ which means there are two degenerate energy levels with opposite parities.
When $N\pmod 8\neq2,6$, either $P$ preserves the parity ( $N\pmod 8=0,4$ )  or no conserved quantity ( $ N $ is odd ),
thus $P^2=+1$ means $Q_q$ belongs to Class AI(GOE) with no degeneracy,
$P^2=-1$ means $Q_q$ belongs to Class AII(GSE) with double degeneracy.
The classification and the energy level degeneracy are
summarized in the 4th and 5th row of \mbox{ Table \ref{tab:SYKq} } respectively

All these results for both even and odd $ N $ have been achieved before in \cite{You2017,hybrid}
in the enlarged Hilbert space by adding an extra fermions at $ \infty $ when $ N $ is odd.
However, here we achieved the same results in the minimal Hilbert space in both even and odd $ N $ in a unified scheme.
Our new method maybe the most powerful and convenient way to achieve new results as shown in the following cases.
%For $N$ odd case, the Hamiltonian have no block structure
%thus the ground state degeneracy is determined by internal degeneracy of RMT ensemble;
%For $N\pmod 8=0,4$, no relation between two different parity blocks
%thus the ground state degeneracy is equal to internal degeneracy of RMT ensemble;
%For $N\pmod 8=2,6$, $P$ relate two different parity blocks
%thus the ground state degeneracy is twice internal degeneracy of RMT ensemble.

\begin{figure}%[!htb]
\includegraphics[width=\linewidth]{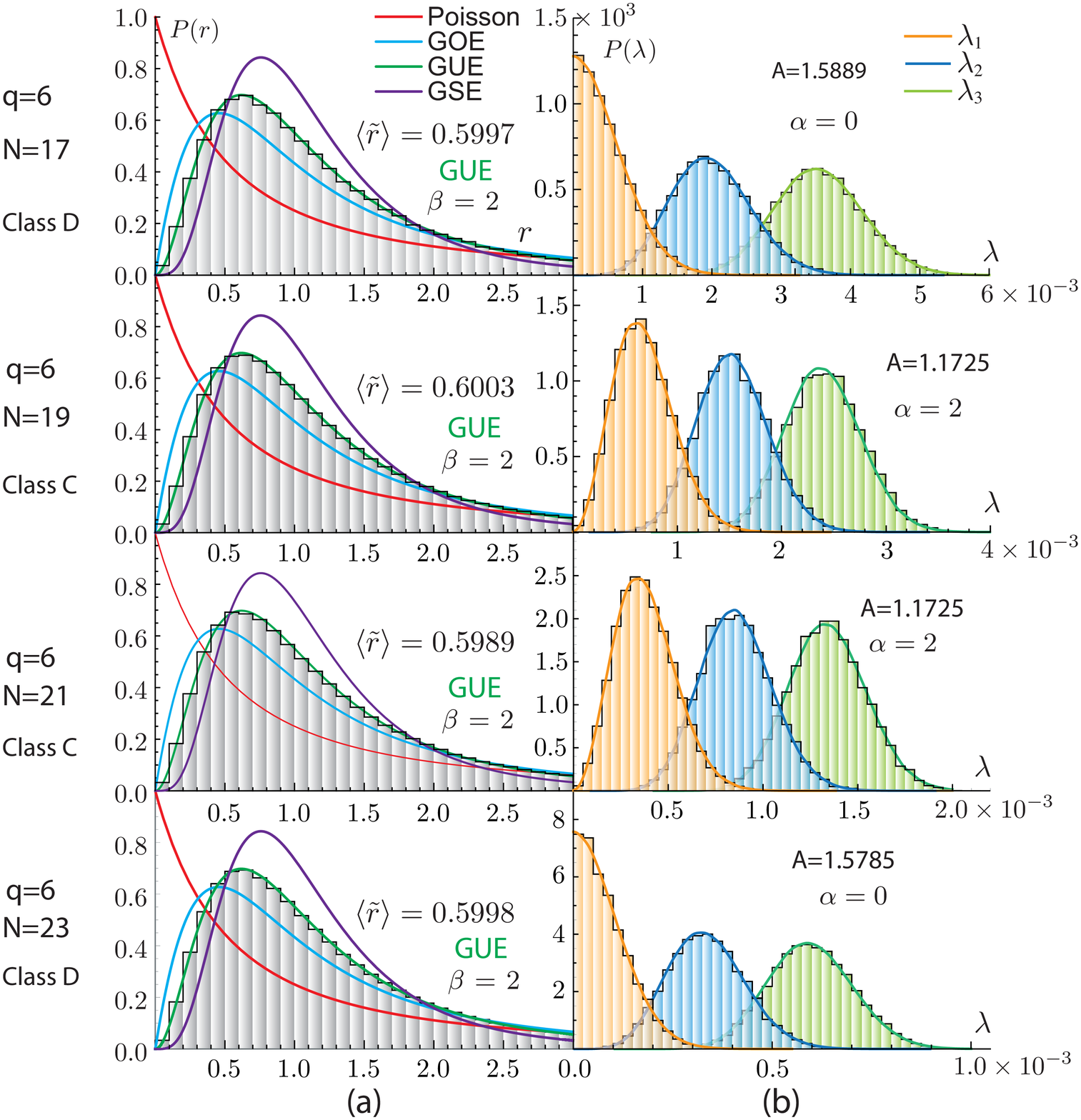}
\caption{
%Many-body energy level statistics of $q=6$ SYK model for
%$N=17,19,21,23$ in (a) bulk and (b) edge.
(a) Distributions of the consecutive level spacing ratio $r$ in $q=6$ SYK model.
%are presented to show level statistics in the bulk.
(b) Distributions of the smallest 3 energy level in $q=6$ SYK model.
%are presented to show level statistics in the hard edge.
The smooth curves for $\lambda_1,\lambda_2,\lambda_3$
are obtained from numerical diagonalization of corresponding random matrix ensembles with size $10^3$
averaged over $10^6$ samples.
The class D and class C have the same bulk index $\beta=2$, but different edge exponent: $\alpha=0$ and $\alpha=2$ respectively.
The $\beta=2$ can be extracted from the distribution $P(r)$
or $\langle\tilde{r}\rangle_W\approx0.6027$.
The $\alpha$ value can be extracted from  distribution $P(\lambda_1)$,
or moment ratio $A=\langle \lambda_1^2\rangle/\langle \lambda_1\rangle^2$.
For class D and C, $\alpha=0,2$ leads to $A_W\approx1.5795, 1.1753$ respectively. }
\label{fig:SYKq6}
\end{figure}

%q=2
{\color{blue}\textit{4. SYK models with $q\pmod 4=2$.}}---
Since $q(1+N)$ is even,
two operators always anti-commute with Hamiltonian $\{P,Q_q\}=\{R,Q_q\}=0$,
and $\Lambda=PR$ is the conserved  parity if $N$ is even
or identity operator if $N$ is odd.
Depending on commutation relation between $P$ and $\Lambda$, the value of $P^2$, one find the following classification:
When $N\pmod 8=2,6$, both operators swap the parity, thus $Q_q$ is in Class A(GUE);
when $N\pmod 8\neq2,6$, either $P$ preserves parity ( $N\pmod 8=0,4$ ) or no conserved quantity ( $ N $ is odd ),
thus $P^2=+1$ means $Q_q$ belongs to Class D(BdG)
and $P^2=-1$ means $Q_q$ belongs to Class C(BdG).
There is no degeneracy in all these cases.
The classification and the energy level degeneracy is
listed in the 6th and 7th row of Table \ref{tab:SYKq} respectively.

%For $N$ odd case, the Hamiltonian has no block structure
%thus the ground state degeneracy is determined by internal degeneracy of RMT ensemble;
%For $N\pmod 8=2,6$, $P$ relate two different parity blocks by a minus sign
%and class A have no mirror symmetry,
%thus the ground state degeneracy is 1;
%For $N\pmod 8=0,4$, no relation between two different parity blocks
%thus the ground state degeneracy is equal to internal degeneracy of RMT ensemble.

The classifications and degeneracy with even $N$
are consistent with those in \cite{Kanazawa2017}.
However, our results with odd $ N $ are new.
We present an ED study of ELS %energy level statistics
for $q=6$ SYK model with $N=17,19,21,23$ in Fig.\ref{fig:SYKq6}.
The random matrix indices $\beta$ and $\alpha$ are extracted from
the probability distribution function $P(r)$ and $P(\lambda_1)$,
or $\langle\tilde{r}\rangle$ and $A$ respectively.
Both methods lead to $(\beta,\alpha)=(2,0),(2,2),(2,2),(2,0)$
for $N=17,19,21,23$ respectively. We also checked the degeneracy is always 1.
These results match the classification and degeneracy listed in Table I.% and II.

\begin{figure}%[!htb]
\includegraphics[width=0.98\linewidth]{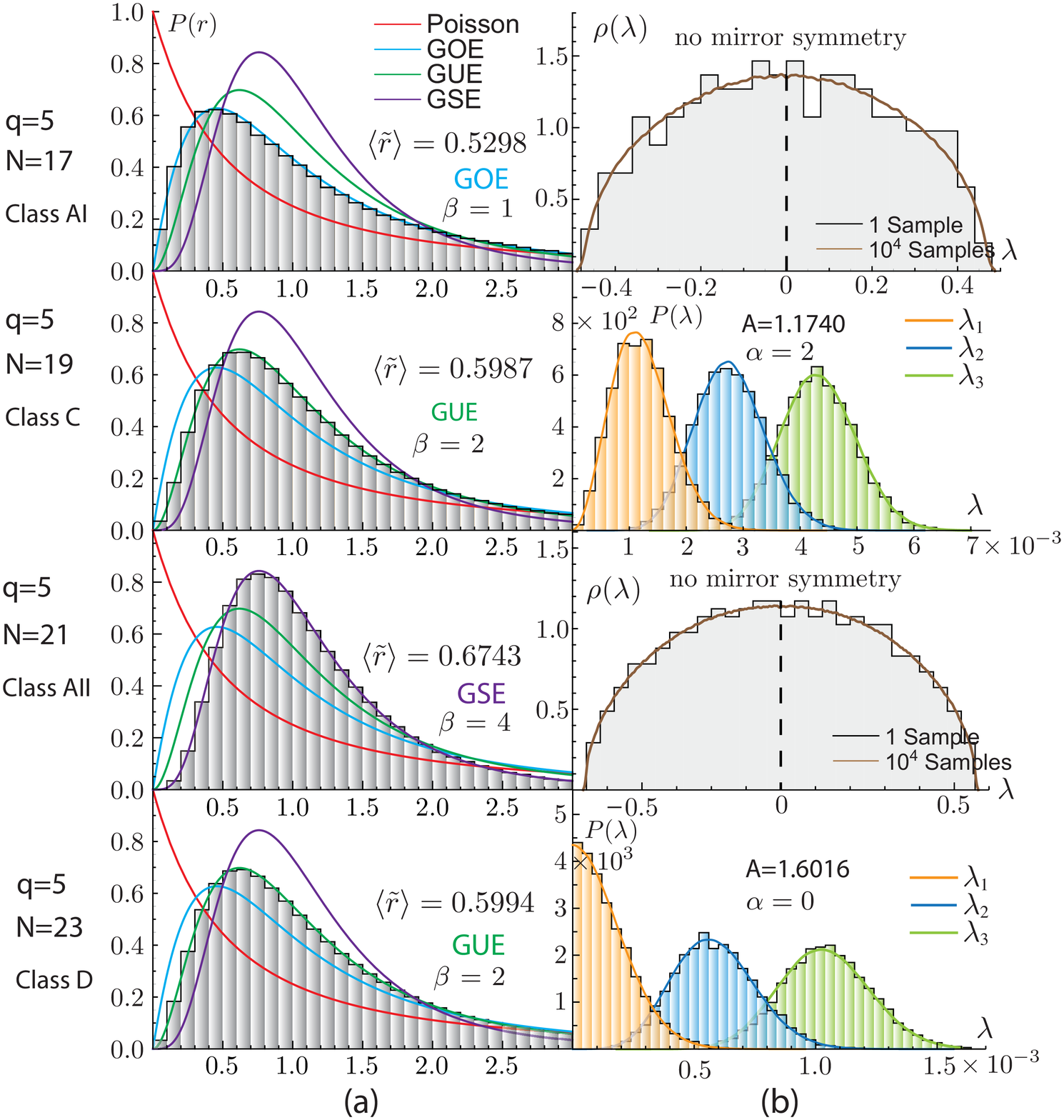}
\caption{
The same notation as Fig. \ref{fig:SYKq6}, but for $q=5$ SYK supercharge.% and $N=17,19,21,23$.
%The spectral density from 1 sample is plotted in (b)
%to show the spectral mirror symmetry is absent
%for a single realization of $C_{i_1,\cdots,i_q}$.
The class AI and AII have $\beta=1,4$,
but no mirror symmetry, thus no well-defined $\alpha$ exponent.
The $\beta=1,4$ can be inferred from $\langle\tilde{r}\rangle_W\approx0.5359,0.6762$ respectively.
For $N=17,21$ cases which are in AI and AII class respectively, the spectral density from one sample is plotted in (b)
to show the spectral mirror symmetry is absent for a single realization of $C_{i_1,\cdots,i_q}$.
But the average over the disorders recovers the RMT semi-circle law \cite{caution}.
%GOE $0.5359$, GUE $0.6027$, GSE $0.6762$
}
\label{fig:SYKq5}
\end{figure}

%q=1
{\color{blue}\textit{5. SYK supercharge with $q\pmod 4=1$.}}---
Now $q(1+N)$ is odd for even $N$ and even for odd $N$,
%so even $N$ and odd $N$ behaviour differently.
%Furthermore, no conserved quantity exist for any $N$.
the former leads to the chiral operator $ \Lambda=(-1)^F $,
the latter leads to $ \Lambda=1 $, so neither will lead to a conserved quantity.
In the following, we discuss even $ N $ and odd $ N $ respectively.

When $N\pmod 8=0,4$, $\{P,Q_q\}=[R,Q_q]=0$.
%thus $P$ ($R$) anti-commutes (commutes) with $Q_q$.
From the values of $P^2$ and $R^2$ in Table \ref{tab:SYKq},
we find: $N\pmod 8=0$ is class BDI(chGOE),
thus $Q_q$ has no degeneracy, but a mirror symmetry,
so $ H=Q_q^2$ has 2-fold degeneracy;
$N\pmod 8=4$ is class CII(chGSE),
thus $Q_q$ has double degeneracy and also a mirror symmetry,
so $ H=Q_q^2$ has 4-fold degeneracy.
When $N\pmod 8=2,6$, $[P,Q_q]=\{R,Q_q\}=0$.
%thus $P$ ($R$)  commutes ( anti-commutes ) with $Q_q$.
From the values of $P^2$ and $R^2$ in Table \ref{tab:SYKq},
we obtain:
$N\pmod 8=2$ is class CI(BdG),
thus $Q_q$ has no degeneracy, but a mirror symmetry,
so $ H= Q_q^2$ has 2-fold degeneracy;
$N\pmod 8=6$ is class DIII(BdG),
thus $Q_q$ has double degeneracy and also a mirror symmetry,
so $ H=Q_q^2$ has 4-fold degeneracy.

When $N\pmod 8=1,5$, $[P,Q_q]=[R,Q_q]=0$.
%thus $P$ and $R$ commute with $Q_q$.
From the values of $P^2$ and $R^2$  in Table \ref{tab:SYKq},
we obtain:
$N\pmod 8=1$ is Class AI(GOE),
thus $Q_q$ has no degeneracy, no mirror symmetry either,
so $H= Q_q^2$ has no degeneracy.
$N\pmod 8=5$ is Class AII(GSE),
thus $ Q_q$ has double degeneracy and no mirror symmetry,
so $H=Q_q^2$ have 2-fold degeneracy.
When $N\pmod 8=3,7$, $\{P,Q_q\}=\{R,Q_q\}=0$.
%thus $P$ and $R$ anti-commute with $Q_q$.
From the values of $P^2$ and $R^2$ in Table \ref{tab:SYKq},
we obtain:
$N\pmod 8=3$ is Class C(BdG),
thus $Q_q$ has no degeneracy, but a mirror symmetry,
so $ H=Q_q^2$ has 2-fold degeneracy.
$N\pmod 8=7$ is Class D(BdG),
thus $Q_q$ has no degeneracy, but a mirror symmetry,
so $ H=Q_q^2$ has 2-fold degeneracy.

The classification for $q\pmod 4=1$ and the energy level degeneracy
are summarized in the 8th, 9th (for $ Q $) and
10th row (for $ H=Q^2 $) of Table \ref{tab:SYKq} respectively.
The results with even $ N $ are consistent with those in \cite{Kanazawa2017}.
However, the results with odd $N$ are new.
We present an ED study of energy level statistics
for $q=5$ SYK supercharge with $N=17,19,21,23$ in Fig.\ref{fig:SYKq5}.
Since the index $\alpha$ is only defined for the cases with mirror symmetry,
when a mirror symmetry is absent,
we plot the spectral density averaged from $1$ sample and $10^4$ samples.
It shows the mirror symmetry is absent for every single realization of $C_{i_1,\cdots,i_q}$,
but still emerges after making disorder averages.
The random matrix indices $\beta$ and $\alpha$, if exist, are extracted from
probability distribution $P(r)$ and $P(\lambda_1)$
or $\langle\tilde{r}\rangle$ and $A$.
Both methods lead to $(\beta,\alpha)=(1,-),(2,2),(4,-),(2,0)$
for $N=17,19,21,23$ respectively. Both the indices $ ( \alpha, \beta) $ and the energy level degeneracy match
our classification and  degeneracy listed in Table I. %and II.

\begin{figure}%[!htb]
\includegraphics[width=0.96\linewidth]{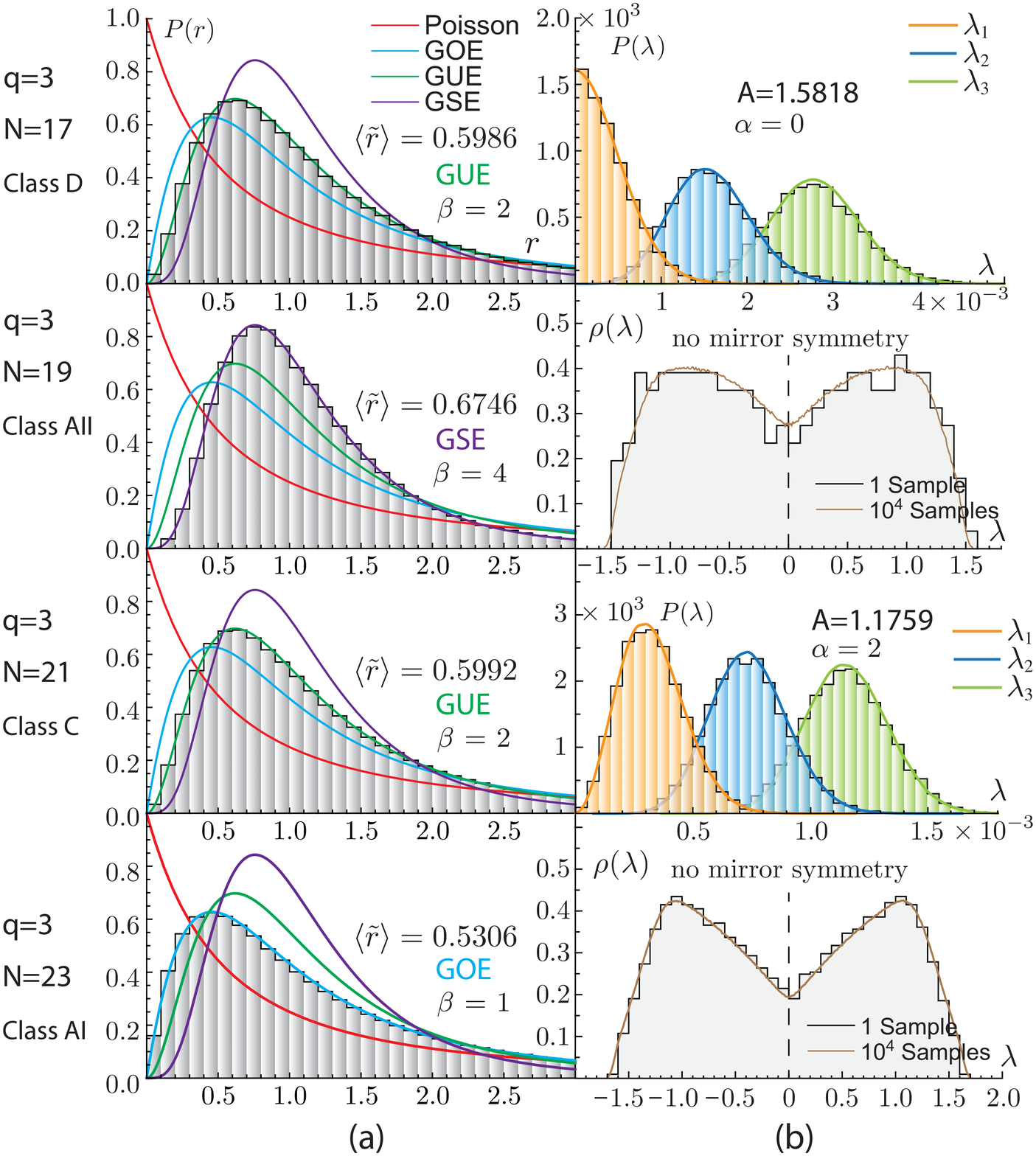}
\caption{The same notation as Fig. \ref{fig:SYKq5} %Many-body energy level statistics of
but for $q=3$ SYK supercharge.
Note that the RMT classes appear in a different order than in the $q\pmod 4=1$ case.
There is also a large dip at $ E=0 $ for the AI, AII class at $ N=23,19 $ respectively.
In fact, the dip always exists for  $ q=3 $ at any $ N \pmod 8 $ ( data not shown ).
However, it disappears and gets back to the semi-circle law for  $ q=7 $ at any $ N \pmod 8 $ ( data not shown ).
The physical origin of this large deviation from the expected semi-circle law only at $ q=3 $ is not known \cite{caution,blowup}.
}
\label{fig:SYKq3}
\end{figure}

%q=3
{\color{blue}\textit{6. SYK supercharge with $q\pmod 4=3$.}}---
Now $q(1+N)$ is odd for even $N$ and even for odd $N$.
The former leads to the chiral operator $ \Lambda=(-1)^F $,
the latter leads to $ \Lambda=1 $, so neither will lead to conserved quantity.
In the following, we discuss even $ N $ and odd $ N $ respectively.

When $N\pmod 8=0,4$, $[P,Q_q]=\{R,Q_q\}=0$,
From the values of $P^2$ and $R^2$ in Table \ref{tab:SYKq},
we obtain:
$N\pmod 8=0$ is class BDI(chGOE),
thus $Q_q$ has no degeneracy, but a mirror symmetry,
so $ H=Q_q^2$ has 2-fold degeneracy;
$N\pmod 8=4$ is class CII(chGSE),
thus $Q_q$ has double degeneracy and also a mirror symmetry,
so $Q_q^2$ has 4-fold degeneracy.
When $N\pmod 8=2,6$, $\{P,Q_q\}=[R,Q_q]=0$.
%thus $P$ ($R$) is anti-commute (commute) with $Q_q$.
From the values of $P^2$ and $R^2$ in Table \ref{tab:SYKq},
we obtain:
$N\pmod 8=2$ is class DIII(BdG),
thus $Q_q$ has double degeneracy and also a mirror symmetry,
so $ H=Q_q^2$ has 4-fold degeneracy;
$N\pmod 8=6$ is class CI(BdG),
thus $Q_q$ has no degeneracy, but a mirror symmetry,
so $Q_q^2$ has 2-fold degeneracy.

When $N\pmod 8=1,5$, $\{P,Q_q\}=\{R,Q_q\}=0$.
%thus both $P$ and $R$ anti-commute with $Q_q$.
From the values of $P^2$ and $R^2$ in Table \ref{tab:SYKq},
we obtain:
$N\pmod 8=1$ is Class D(BdG),
thus $Q_q$ has no degeneracy, but a mirror symmetry,
so $H=Q_q^2$ has 2-fold degeneracy;
$N\pmod 8=5$ is Class C(BdG),
thus $Q_q$ has no degeneracy, but a mirror symmetry,
so $Q_q^2$ have 2-fold degeneracy.
When $N\pmod 8=3,7$, $[P,Q_q]=[R,Q_q]=0$.
%thus both $P$ and $R$ commute with $Q_q$.
From the values of $P^2$ and $R^2$ in Table \ref{tab:SYKq},
we obtain:
$N\pmod 8=3$ is Class AII(GSE),
thus $Q_q$ has double degeneracy, but no mirror symmetry,
so $ H=Q_q^2$ has 2-fold degeneracy;
$N\pmod 8=7$ is Class AI(GOE),
thus $Q_q$ has no degeneracy and no mirror symmetry either,
so $ H=Q_q^2$ have no degeneracy.

The classification for $q\pmod 4=3$ and  the energy level degeneracy are summarized in the 11th, 12th and 13th row of Table \ref{tab:SYKq}
 respectively.
Classifications and degeneracy with even $N$ are consistent with \cite{Kanazawa2017}.
%Notice the degeneracy for odd $N$ is only half Ref. \cite{Fu2017}'s result,
%because his definition of degeneracy of odd $N$ case is in a Hilbert space
%including an extra decoupled Majorana fermion, which is twice bigger than our's.
However, or results with odd $N$ are new. We
present an ED study of $q=3$ SYK supercharge for $N=17,19,21,23$
in Fig.\ref{fig:SYKq3}.
These data show $(\beta,\alpha)=(2,0),(4,-),(2,2),(1,-),$
for $N=17,19,21,23$ respectively. Both the indices $ ( \alpha, \beta) $ and the energy level degeneracy match
our classification and  degeneracy listed in Table I. %and II.

%{\color{blue}\textit{Comparing Random Matrix Hard Edges and SYK Large $N$ limit.}}---

{\color{blue}\textit{7. Discussion and Summary.}}---
In the $ 1/N $ expansion of the SYK model \cite{Polchinski2016,Maldacena2016,Gross2017,Qi} and SUSY SYK model \cite{Fu2017} with any $q$-body
interaction, a $1/q$  expansion follows which can be used to derive 2- or 4-point functions at any $ \beta J $ scale.
At a double scaling limit $ N \rightarrow \infty$, $q \rightarrow \infty $, but keeps $ q^2/N $ at a fixed ratio,
one may map the effective action to a Liouville conformal field theory \cite{Bagrets2016,liu1,liu2,Cotler2017}.
Here, we put SYK with even $ q $ and SUSY SYK with odd $ q $,
$ N $ even or odd on the same footing and find the $ N \pmod 8 $ and  $q \pmod 4=0$ double Bott periodicity.
Unfortunately, the double Bott periodicity can not be seen in the $1/ N$, $1/q $ expansion or the double scaling limit.
This maybe due to the huge time scale separations between the $1/N $ expansion method and the RMT classifications outlined in the introduction.
Indeed, the Lyapunov exponent can  be extracted by evaluating  out of time ordered correlation (OTOC) function
in the large $ N $ limit where there is a wide separation between the dissipation time $ t_d \sim 1/\beta $ and the
scrambling time $ t_s\sim t_d \log N $.
At any finite size $ N=23 $ in Fig.1-3, this window may be still to narrow to extract the  Lyapunov exponent \cite{Fu2016}.
Non-perturbative effects maybe involved  to see the  double Bott periodicity  in the $ 1/N $ expansion.
The reason why one can put SYK  and SUSY SYK in the same periodic table maybe also due to the fact that the $ {\cal N}=1 $ SUSY
is broken at any finite $ N $, but only non-perturbatively.  Indeed,
the ground state energy is lifted from zero to $ e^{-N} $ which is in accessible to the perturbative $ 1/N $ expansion.
So one  expect this SUSY breaking at any finite $ N $ is a non-perturbative instanton tunneling  effects.
Similarly, in the perturbative $1/N$ expansion of the
Dicke model \cite{KAM} in the super-radiant phase, one must consider the non-perturbative instanton tunneling events  to find the $ e^{-N} $ splitting between the two states with opposite parities.

It is interesting to make an analogy to the periodic tables of TI and TSC first put forward in \cite{Kitaev2009,four}.
Despite using the same Cartan labels and same sets of anti-unitary operators, it is on topological equivalent classes
in Bott periodicity in the space dimension $ d \pmod 8 $. It is also for a gapped bulk states with gapless edge modes.
While ours is for gapless quantum spin liquids whose quantum chaos can be characterized by bulk energy level statistics 
and edge exponents.
Later, adding more symmetries such as translational symmetries, point group symmetrise or non-isomorphic symmetries
lead to more rich periodic tables for topological crystalline symmetries \cite{tenfold}.
However, for  many body interacting electron systems, the anti-unitary operators are not enough,
the SPT or SET phases maybe classified by using more
advanced mathematical tools such as co-homology, cobordism and tensor categories \cite{tenfold,wen} which were already used
in rational conformal field theory (RCFT) and also topological quantum computing.

%which make classification as simple as possible.
The most compact, systematic and complete classification method developed in this work can be applied to
all kinds of generalizations of Majorana SYK models such as the Brownian SYK \cite{Brown},
the colored-SYK model \cite{Gross2017,color}, two indices SYK model \cite{Ye}
or other Majorana SYK-like model with global $ O(M), U(M) $ symmetry \cite{Qi,Yoon}.
This is just putting more symmetries lead to more conserved quantities, also more operators, therefore generating more
complicated periodic tables ( for the color degree of freedoms, see \cite{hybrid},
for $ U(1) $ complex SYK or $ {\cal N}=1 $ SUSY SYK, see SM section E ).
Its advantages over the extended schemes ( see SM ) may be more evident and dramatic as the periodic tables get more rich and complicated.
It may also be applied to study quantum chaos in the colored (Gurau-Witten) Tensor model \cite{tensor1,tensor2}
or uncolored  Klebanov-Tarnopolsky tensor model \cite{tensorrev}.
A new moment ratio $A$ proposed in this work is a powerful and efficient tool to identify the edge exponent in the 3
chiral Classes and 4 BdG Classes.
It may also be used to characterize the edge transition
between different random matrix ensembles, such as the edge transitions in Hybrid $ {\cal N}=1 $ SUSY SYK model \cite{un}.

The $ q=4 $ SYK models may be experimentally realized in various cold atom and solid state systems \cite{cold,solid}.
%These include some of the standard platforms for generating Majorana zero modes
%in prosimitized quantum wires and topological insulator- superconductor interfaces
%as well as electrons in the lowest Landau level in a graphene flake with an irregular boundary.
%So the different quantum chaos in the periodic table may be observed in the future experiments.
The periodic table tells that an isolated $ q=4 $ SYK island should show
sensitive even/odd effects in the number of particles inside the island which is an interesting mesoscopic quantum phenomena.
% which can be detected experimentally.
Due to its quantum chaotic nature, the SYK islands always satisfy
the Eigenstate Thermalization Hypothesis (ETH), therefore is an equilibrium system with a well defined temperature ( so is a
black hole ).
The number of particles $ N $ in an isolated island can be easily adjusted, so the $N\pmod 8$ Bott periodicity  may be
observed experimentally. However, it is experimentally more challenging to adjust $q$-body interaction to observe $q \pmod 4$
Bott periodicity. The Coulomb interaction in a solid system  automatically leads to $ q=4 $ body interaction.
The $ q=6 $ body interaction and the SUSY SYK with an odd $q$-body such as $ q=3 $ interaction  maybe realized in cold atom systems.

The classical black holes can be classified just by 3 classical quantities the mass $ M $, charge $ Q $ and the angular momentum $ L$ ( No hair theorem ).
So far, the charge neutral$ Q=0 $ quantum black holes are only discussed in the AI (GUE) class \cite{Cotler2017,Brown,genus}.
It remains unknown if  the periodic Table \ref{tab:SYKq} can be applied to the JT gravity side \cite{JT-1,JT-2} and its genus expansion \cite{genus}.
It remains to be seen if it also implies the classifications of quantum natures of charge neutral black holes.

%We think understanding the full periodic pattern of SYK models
%might be helpful to making connection between
%quantum chaos in early time and late time.

%{\color{blue}\textit{Acknowledgements.}}---
We thank S Shenker for helpful discussions. We acknowledge AFOSR FA9550-16-1-0412 for supports.
This research at KITP was supported in part
by the National Science Foundation under Grant
No. NSF PHY-1748958.

\bibliographystyle{apsrev4-1}
%\bibliography{mybib}{}

%=======================================================
%----		supplemental materials		--------
%=======================================================
%%%%%%%%%% Merge with supplemental materials %%%%%%%%%%
%\onecolumngrid
%\pagebreak
\widetext
\begin{center}
\textbf{\large Supplementary Materials for
`` Periodic Table of SYK and supersymmetric SYK models ''}
\end{center}
%%%%%%%%%% Merge with supplemental materials %%%%%%%%%%
%%%%%%%%%% Prefix a "S" to all equations, figures, tables and reset the counter %%%%%%%%%%
\setcounter{equation}{0}
\setcounter{figure}{0}
\setcounter{table}{0}
\setcounter{page}{1}
\makeatletter
\renewcommand{\theequation}{S\arabic{equation}}
\renewcommand{\thefigure}{S\arabic{figure}}
\renewcommand{\thetable}{S\arabic{table}}
\renewcommand{\bibnumfmt}[1]{[S#1]}
%\renewcommand{\citenumfont}[1]{S#1}
%%%%%%%%%% Prefix a "S" to all equations, figures, tables and reset the counter %%%%%%%%%%

In the supplementary materials,
(A) we first review some known results on 10-fold way random matrix classifications, bulk energy level statistic
characterized by the ratio of nearest neighbour spacing and next nearest neighbour spacing, then
we propose a new moment ratio $A$ to measure the edge exponent $\alpha$ effectively.
(B)  we work out explicitly the fine structure of the energy levels using
the two anti-unitary operators in the minimal Hilbert space.
(C)When $N$ is odd, we present an independent classification of SYK and SUSY SYK models and count the energy level degeneracy
by adding a decoupled Majorana into system .
In the SUSY SYK with odd $ N $, we find a new conserved parity $ (-1)^F i \chi_{\infty} $
which commutes with the super-charge $ Q_q $ and do the classification
and count the degeneracy in the super-charge $ ( Q_q ,( -1)^F i \chi_{\infty} ) $ basis.
(D) For the SUSY SYK  ( odd $q $ ), for even $ N $, we do the classification and count the energy level degeneracy
in the $ (-1)^F $ basis, for odd $ N $, by adding a decoupled Majorana into system, we
find a new conserved parity  $ i \chi_{\infty} Q_q $ which commutes with $ (-1)^F $ and do the classification
and count the degeneracy in the $ ( (-1)^F,  i \chi_{\infty} Q_q  )  $ basis.
(E) Finally We also develop a systematic classification of the most general complex SYK models.

It is constructive to contrast different classification schemes in (B), (C) and (D) which also count the energy level degeneracy
from different perspectives. The comparison establishes interesting and also quite intricate
 connections among different classification schemes. It
may also  stress the compactness and effectiveness of the minimum scheme developed in the main text
where one has a minimum number of complete anti-unitary or unitary symmetry operators and conserved quantities.
It also provides additional insights on the global picture of the periodic table.
%As shown in the main text and the SM, different classification schemes can be developed to
%classify, but the most compact and concise way is to do it  in a minima Hilbert space where we

\section{ A. Ten-fold way, bulk energy level statistics and edge exponents }

%Fine tuning the table
\renewcommand{\arraystretch}{1.4}%Default value is 1
\renewcommand\tabcolsep{6pt}%Default value is 0pt

\begin{table}[!htb]
\caption{ Ten fold way of RMT classes, adapted from \cite{AZ,Kitaev2009,four,kane,tenfold}.
$T_+$ ($T_-$) denotes an anti-unitary operator that commutes (anti-commutes) with the Hamiltonian,
$\Lambda= T_+ T_- $ is an unitary operator that anti-commutes with the Hamiltonian.
The symbol --- means such an  operator does not exist.
In the presence of conserved quantities, $T_+$ ($T_-$) need also commute with all the compatible conserved quantities. }
%Adapted from JHEP09(2017)050.

\begin{tabular}{ c|cccccccccc }
\toprule
RMT Class   &A	    &AI	    &AII    &AIII   &BDI    &CII    &C	    &CI	    &D	    &DIII   \\
\hline
$T_+^2$	    &---    &$+1$   &$-1$   &---    &$+1$   &$-1$   &---    &$+1$   &---    &$-1$   \\
$T_-^2$	    &---    &---    &---    &---    &$+1$   &$-1$   &$-1$   &$-1$   &$+1$   &$+1$   \\ \hline
$\Lambda^2$ &---    &---    &---    &$1$    &$1$    &$1$    &---    &$1$    &---    &$1$   \\ \hline
%\toprule
\end{tabular}
\label{tab:RMT}
\end{table}

In the pioneering works \cite{Wigner,Dyson} , Wigner and Dyson classified
3 symmetry classes by
an anti-unitary operator $T_+$, i.e. time-reversal symmetry operator,
which commutes with Hamiltonian.
In the absence of $T_+$, the symmetry class is A(GUE);
while $T_+^2=+1$, it is AI(GOE)
and $T_+^2=-1$, it is AII(GSE).
These 3 symmetry classes also known as 3 Wigner-Dyson classes.
Later, 7 additional symmetry classes were found by
considering another anti-unitary operator $T_-$ and an unitary operator $\Lambda$
which anti-commutes with Hamiltonian.
All the 10 symmetry classes \cite{AZ} can be completely determined
by squared values of $T_+^2$, $T_-^2$ and $\Lambda^2$.
Dropping the $T_-$ and $\Lambda$ operators recovers the original 3 Wigner-Dyson classes.
The symmetry classification scheme also applies to topological insulators
and superconductors \cite{Kitaev2009,four,tenfold} in the context of topological equivalent classes.
To be self-contained, we list the 10 fold way of RMT in the Table \ref{tab:RMT} which was heavily consulted in the main text
and in the SM.

%This relation is explicitly given in Table 1 of Ref.\cite{Kanazawa2017}.
In the presence of conserved quantities \cite{color}, one need to identify all the anti-unitary operators which preserve
all the compatible conserved quantities.
The classification need be applied to the each block of the Hamiltonian.

\bigskip
{\bf A1. Bulk Energy level statistics and bulk index $ \beta $. }
\bigskip

For the 3 Wigner-Dyson classes, %: A(GUE), AI(GOE), AII(GSE).
no mirror symmetry exists in the energy levels, the joint probability distribution for all the eigen-values can be described by
$P(\{\lambda_i\})\propto\prod_{i<j}|\lambda_i-\lambda_j|^\beta
\prod_n e^{-\lambda_n^2}$,
where $\beta$ is the Wigner-Dyson index characterizing the strength of level repulsion.
For class A(GUE), AI(GOE), AII(GSE), $\beta=2,1,4$ respectively.

Among the 7 additional classes: AIII(chGUE), BDI(chGOE), CII(chGSE),
C(BdG), D(BdG), CI(BdG), DIII(BdG) which owns a mirror symmetry,
the first three are chiral ensembles relevant to the systems with Dirac fermions such as QCD;
the last four are BdG class, arising via Bogolioubov-de Gennes (BdG) mean field approximation of superconductors.
Due to the mirror symmetry, every energy level appears in pair $ \pm \lambda $  and
the joint probability density distributions for all the eigen-values can be described by
$P(\{\lambda_i\})\propto\prod_{i<j}|\lambda_i^2-\lambda_j^2|^\beta
\prod_n |\lambda_n|^\alpha e^{-\lambda_n^2}$,
where $\alpha$ is the edge exponent characterizing the repulsion from ``hard edge'' at $\lambda=0$.
(On the contrary, the soft edge is referred to the spectral edge at the maxima $|\lambda|$.)
The $\alpha$ index of the 7  classes with the mirror symmetry is listed in Table \ref{tab:RMTalpha} .
Note that the $\alpha$ index is not defined for the 3 original Wigner-Dyson classes due to the lacking of mirror symmetric,
thus no special point in the spectrum after unfolding procedure.
%$\beta$ from bulk

In performing exact Diagonizations (ED), unfolding procedure is needed when comparing numerical data with RMT predictions,
so it may not be convenient to directly use $P(\{\lambda_i\})$. To avoid this kind of procedures,
a universal ratio $r_n=(\lambda_{n+2}-\lambda_{n+1})/(\lambda_{n+1}-\lambda_n)$
of the consecutive energy level spacings   can be constructed
to describe the \emph{bulk} energy level statistics. Due to the cancelation on the dependence
on the density of states in the ratio, the unfolding procedure can be avoided.
By performing the exact calculation for $n=3$ case, the authors in Ref.\cite{Atas2013} derived
a Wigner-like surmises for the distribution of $r_n $:
$P_W(r)\propto(r+r^2)^\beta/(1+r+r^2)^{1+3\beta/2}$.
The level repulsion is reflected in the asymptotic behaviour $P(r)\sim r^{\beta}$ at small $r$.
The most efficient way to extract $\beta$ index is to study the expectation $\langle r\rangle$ or
its cousin $\langle \tilde{r}\rangle$ with $\tilde{r}_n=\min(r_n,1/r_n)$.
It was documented that $\langle\tilde{r}\rangle=0.5359,0.6027,0.6762$ for $\beta=1,2,4$ respectively.
Thus, $\langle\tilde{r}\rangle$-parameter was also used to
determine transitions between random matrix ensembles with different $\beta$ indexes
\cite{hybrid}. However, in the presence of energy level degeneracy in the same conserved quantity sector, the ratio of next nearest neighbor energy level spacing $r^{\prime}_n=(\lambda_{n+2}-\lambda_{n})/(\lambda_{n}-\lambda_{n-2} )$, its
expectation value $\langle r^{\prime}\rangle$ or
its cousin $\langle \tilde{r}^{\prime}\rangle$ with $\tilde{r}^{\prime}_n=\min(r^{\prime}_n,1/r^{\prime}_n ) $
must be used to characterize the transitions. The values of $\langle r^{\prime}\rangle$ and $\langle \tilde{r}^{\prime}\rangle$
for $\beta=1,2,4$ are documented in \cite{hybrid}.
%or Chaotical to non chaodic transiton.
%[cite hybrid papers]

\bigskip
{\bf A2. Wigner surmises and a new moment ratio for the 7 classes with a mirror symmetry  to determine the edge exponent $\alpha$. }
\bigskip

For the seven classes with a mirror symmetry, in addition to the bulk energy level statistics,
one may also need to focus on the universality near the hard edge $\lambda=0$.
One way is to integrate out all the energy levels in
$P(\{\lambda_i\})$ except the smallest positive one
which result in the probability distribution function of the smallest positive eigenvalues $P_1(\lambda)$.
The probability distribution function of
$k$-th positive eigenvalues $P_k(\lambda)$ can be similarly obtained.
Exact results of $P_k(\lambda)$ were known for the three chiral classes describing quarks in QCD
\cite{rmt-1,rmt-2}. Here, we only list the exact results of distribution function of the smallest positive eigenvalue
(denotes as $\lambda$) for the three chiral classes, AIII(chGUE), BDI(chGOE) and CII(chGSE):
\begin{align}
    P_\text{AIII}(\lambda)
	&=(\lambda/2)e^{-\lambda^2/4} \label{eq:AIII}\\
    P_\text{BDI}(\lambda)
	&=[(2+\lambda)e^{-\lambda^2/8-\lambda/2}]/4 \label{eq:BDI}\\
    P_\text{CII}(\lambda)
	&=(\pi/2)^{1/2}\lambda^{3/2}e^{-\lambda^2/2}I_{3/2}(\lambda) \label{eq:CII}
\end{align}
where $I_n(\lambda)$ is the Modified Bessel functions of the first kind.

%An alternative way is using microscopic spectral density
%$\rho_M(\lambda)=\Delta \rho(\lambda/\Delta)$,
%where $\rho(\lambda)=(1/n)\langle\sum_i^n\delta(\lambda-\lambda_i)\rangle$ is macroscopic spectral density
%and $\Delta$ is the level spacing near $\lambda=0$.
%Analytical expressions of $\rho_M$ are known for all random matrix classes.
%However, due to the relation $\rho_M(\lambda)=\sum_{k=1}^{\infty}P_k(\lambda)$,
%thus first a few $P_k(\lambda)$ will be enough to
%extract the universality near the hard edge.

%Fine tuning the table
%Setting for main text
%\renewcommand{\arraystretch}{1.2}%Default value is 1
%\renewcommand\tabcolsep{2pt}%Default value is 0pt

%Setting for SM
\renewcommand{\arraystretch}{1.4}%Default value is 1
\renewcommand\tabcolsep{6pt}%Default value is 0pt

\begin{table}[!htb]
\caption{
RMT indices $(\alpha, \beta) $ and the corresponding moment ratio $A$ for the 7 classes with a mirror symmetry
(but no topological zero mode).
$A_\text{W}$ is that calculated from the exact $ P_1 (\lambda ) $ of the 3 chiral Classes or our surmise of the 4 BdG classes.
$A_\text{num}$ is that calculated from the exact diagonization of corresponding random matrices of size $N=10^3$
with Gaussian distributed entries, averaged over $ 10^5 $ samples. }
\begin{tabular}{ c|c|c|c|c|c|c|c }
	\toprule%\hline
	RMT $\alpha$
	    &\multicolumn{2}{c|}{0}
	     &\multicolumn{3}{c|}{1}
	      &2 &3\\ \hline
	RMT $\beta$	&1    &2  &1   &2     &4     &2  &4\\ \hline
	Class	&BDI &D &CI &AIII &DIII &C &CII\\ \hline
	$A_\text{W}$
	    &1.6018	&1.5795			%BDI D
	    &1.2732 	&1.2732	    &1.2732	%CI AIII DIII
	    &1.1753				%C
	    &1.1291	\\  			%CII
	\hline
	$A_\text{num}$
	    &1.6021	&1.5832
	    &1.2745 	&1.2738	    &1.2719
	    &1.1735
	    &1.1296	\\
%	\hline
%	$T_+^2$		&$+1$	&---    &$+1$	&---	&$-1$	&---    &$-1$\\
%	$T_-^2$		&$+1$	&$+1$	&$-1$	&---	&$+1$	&$-1$	&$-1$\\
%	$\Lambda^2$	&1	&---    &1	&1	&1	&---    &1\\
	\toprule
\end{tabular}
\label{tab:RMTalpha}
\end{table}

%Here, instead of deriving the exact results for the 4 BdG classes, we will develop
%a Wigner-like surmise which can be used to determine the edge exponent $ \alpha $ very effectively.

However, the exact results of distribution function of the smallest positive eigenvalue
of 4 BdG classes, C, D, CI, and DIII, are not known yet.
Here, instead of deriving the exact results,
we propose Wigner surmises for the distribution function of the 4 BdG class.
This method was proved to be very accurate for
Hermitian and non-Hermitian chiral random matrices ensemble to describe QCD \cite{WS}.
Here, we will show it can be used to determine the edge exponent $ \alpha $ very effectively also for the 4 BdG classes.

Since $P_1(\lambda)$ describes the energy level closest to the hard edge, it shows universal RMT behaviours.
Other $P_k(\lambda), k > 1 $ can be treated in a similar way,
but it is believed that $P_{k>1}$ is less universal than $P_1(\lambda)$.
This is also reflected in the fact that $P_{k}$ with large $k$ may show large deviations when compared with numerical data.
So in the following, we will mainly focus on $P_1(\lambda)$.
Since $\lambda$ has dimension of energy, thus a rescaling $P_1(\lambda)\to cP_1(c\lambda)$
may be needed when comparing with numerical data.
As shown below, typically, one choose
$c\int d\lambda \lambda P_1(\lambda)=\langle\lambda_\text{min}\rangle$
to fix the rescaling factor which need to be the same for all $P_k$.

Starting from the joint probability density for the energy levels with $n=2$
\begin{align}
    P(\lambda_1,\lambda_2)\propto
	|\lambda_1^2-\lambda_2^2|^\beta
	|\lambda_1\lambda_2|^\alpha e^{-(\lambda_1^2+\lambda_2^2)}
\end{align}
where $\propto$ means the normalization constant is ignored.
The distribution function of smallest positive eigenvalue
can be obtained from
\begin{align}
    P_1(\lambda)=\int_{\lambda}^{\infty}d\lambda_2 P(\lambda,\lambda_2)
\end{align}
which can be done analytically.

For BdG class D, we need choose $\beta=2$, $\alpha=0$ and obtain
\begin{align}
    P_\text{D}(\lambda)
	=\frac{1}{\pi}e^{-2\lambda^2}
	[6\lambda-4\lambda^3
	+\sqrt{\pi}e^{\lambda^2}\erfc(\lambda)(3-4\lambda^2+4\lambda^4)]
\label{eq:D}
\end{align}
where $\erfc(\lambda)$ is the complementary error function.
%and $\lim_{s\to\infty}se^{s^2}\erfc(s)=1/\sqrt{2}$
%tells the probability is never divergent.
In order to use $P_\text{D}(\lambda)$ to compare with numerical data,
we need to rescale $P_\text{D}(\lambda)\to \tilde{P}(\tilde{\lambda})=cP_\text{D}(c\lambda)$
such that  $\int_0^\infty d\tilde{\lambda} \tilde{\lambda} \tilde{P}(\tilde{\lambda})
=\langle\lambda_{\min}\rangle$ with $\tilde{\lambda}=c\lambda$.
This requirement sets  the scale factor to be
$c=\frac{7-4\sqrt{2}}{2\sqrt{\pi}\langle\lambda_{\min}\rangle}$ where $\langle\lambda_{\min}\rangle$ is the mean value
of the smallest positive eigenvalue from the numerical data.

 For BdG class C, we need choose $\beta=2$, $\alpha=2$ and obtain
\begin{align}
    P_\text{C}(\lambda)
	=\frac{2}{3\pi}e^{-2\lambda^2}\lambda^2
	[30\lambda-4\lambda^3
	+\sqrt{\pi}e^{\lambda^2}\erfc(\lambda)(15-12\lambda^2+4\lambda^4)]
\label{eq:C}
\end{align}
In order to use $P_\text{C}(\lambda)$ to compare with numerical data,
we need again to rescale $P_\text{C}(\lambda)\to \tilde{P}(\lambda)=cP_\text{C}(c\lambda)$
where the scale factor is found to be
$c=\frac{5(2-\sqrt{2})}{2\sqrt{\pi}\langle\lambda_{\min}\rangle}$.
%and $\langle\lambda_{\min}\rangle$ is mean value
%of the smallest positive eigenvalue from numerical data.

For BdG class DIII, we need choose $\beta=4$, $\alpha=1$ and obtain
\begin{align}
    P_\text{DIII}(\lambda)=4\lambda e^{-2\lambda^2}
\label{eq:DIII}
\end{align}
In order to use $P_\text{DIII}(\lambda)$ to compare with numerical data,
we need rescale $P_\text{DIII}(\lambda)\to \tilde{P}(\lambda)=cP_\text{DIII}(c\lambda)$
where the scale factor is found to be
$c=\frac{\sqrt{\pi}}{2\sqrt{2}\langle\lambda_{\min}\rangle}$.

For BdG class CI, we need choose $\beta=1$, $\alpha=1$ and obtain
\begin{align}
    P_\text{CI}(\lambda)=4\lambda e^{-2\lambda^2}
\label{eq:CI}
\end{align}
which is the same as class DIII.

In fact, the independence of  $P_1(\lambda)$ at $\alpha=1$  on $\beta$ index
{\sl upto to a rescaling factor } can be understood by a direct evaluation of the integral:
\begin{align}
	P_1(\lambda_1)
	=\frac{1}{(N-1)!}
	\Big(\prod_{n=2}\int_{\lambda_1}^\infty d\lambda_n\Big) P(\{\lambda\})
	\propto
	\Big(\prod_{n=2}\int_{\lambda_1}^\infty d\lambda_n\Big)
	\Big(\prod_{i<j}|\lambda_i^2-\lambda_j^2|^\beta\Big)
	\Big(\prod_{n=1} \lambda_n e^{-\lambda_n^2}\Big)
\end{align}

 A change of variable and shift the integral domain lead to:
\begin{align}
    P_1(\lambda_1)
	&\propto
	\lambda_1 e^{-\lambda_1^2}
	\Big(\prod_{n=2}\int_{\lambda_1^2}^\infty
	    |\lambda_n^2-\lambda_1^2|^\beta e^{-\lambda_n^2} d\lambda_n^2\Big)
	\Big(\prod_{1<i<j}|\lambda_i^2-\lambda_j^2|^\beta\Big) \nonumber\\
	&=
	\lambda_1 e^{-N\lambda_1^2}
	\Big(\prod_{n=2}\int_{0}^\infty |\lambda_n^2|^\beta
	e^{-\lambda_n^2} d\lambda_n^2\Big)
	\Big(\prod_{1<i<j}|\lambda_i^2-\lambda_j^2|^\beta\Big)
	=C_{N,\beta}\lambda_1 e^{-N\lambda_1^2}
\end{align}

So it is obvious no $\beta$ dependence in the final result
after a normalization and a suitable rescale of $\lambda_1$
( For example, in the $N\to\infty$, one just need to scale $\lambda\to c\lambda/\sqrt{N}$ ).
If one choose a suitable scale,
all $\alpha=1$ ensemble will have the same $P_1$ as Class AIII.
A rescale of $\lambda_1$ will not affect the moment ratio $A$ defined in Eq.\ref{momentratio},
so $\alpha=1$ will lead to the same $A_W$ in Table \ref{tab:RMTalpha}.

Using the fact $\erfc(\lambda)=1-2\lambda/\sqrt{\pi}+O(\lambda^2)$,
one can check all of these results have the asymptotic behaviour
$P(\lambda)\sim\lambda^\alpha$ when $\lambda$ is small.
In principle, the edge exponent $\alpha$ can be obtained by studying the small $\lambda$ behaviour of $P(\lambda)$,
But in practice, although this method can be used to easily distinguish between $\alpha=0$ and $\alpha>0$ cases, it
may not be efficiently used to tell the differences among $\alpha=1,2,3$ cases.
Inspired by the $r$-parameter introduced to get rid of the dependence on the local density of state and
 determine the bulk energy level statistics efficiently,
we find it is more convenient to introduce  a new  moment ratio  $A$ which is a dimensionless quantity to get rid of the
dependence on the rescaling factor:
\begin{align}
    A=\frac{\langle\lambda_\text{min}^2\rangle}
	   {\langle\lambda_\text{min}\rangle^2}
    =\frac{\int_0^\infty d\lambda \lambda^2 P(\lambda)}
	{\Big[\int_0^\infty d\lambda \lambda P(\lambda)\Big]^2}
	   \geq1
\label{momentratio}
\end{align}
which is obviously indeed independent of the rescaling factor.

%Different $\alpha$ will result in different $A$ value (see Table \ref{tab:RMTalpha}),
%thus $A$ value can be used to identify $\alpha$ index.
%\cite{Atas2013,Akemann2009}

Since the Majorana SYK model do not contain any topological protected zero modes,
there is a one-to-one correspondence  between RMT indices $ ( \beta,\alpha ) $ and RMT classes.
We also perform EDs on random matrices of size $10^3$
from these 7 symmetry classes.
The moment ratio $A$ calculated from the analytical functions for the 7 classes
%\cref{eq:AIII,eq:BDI,eq:CII,eq:D,eq:C,eq:DIII,eq:CI}
\ref{eq:AIII},\ref{eq:BDI},\ref{eq:CII},\ref{eq:D},\ref{eq:C},\ref{eq:DIII},\ref{eq:CI}
and  the numerical data are listed in Table \ref{tab:RMTalpha}.
A comparison between our Winger-like surmise
and numerical data is shown in Fig. S1.
The combination of  two dimensionless $\tilde{r}$ and $A$-values can also be used  to determine the transition
between random matrix ensembles with different $\beta, \alpha$ indices such as that in the hybrid $ {\cal N}=1 $ SUSY SYK model \cite{un}.

\begin{figure}[!htb]
\includegraphics[width=\linewidth]{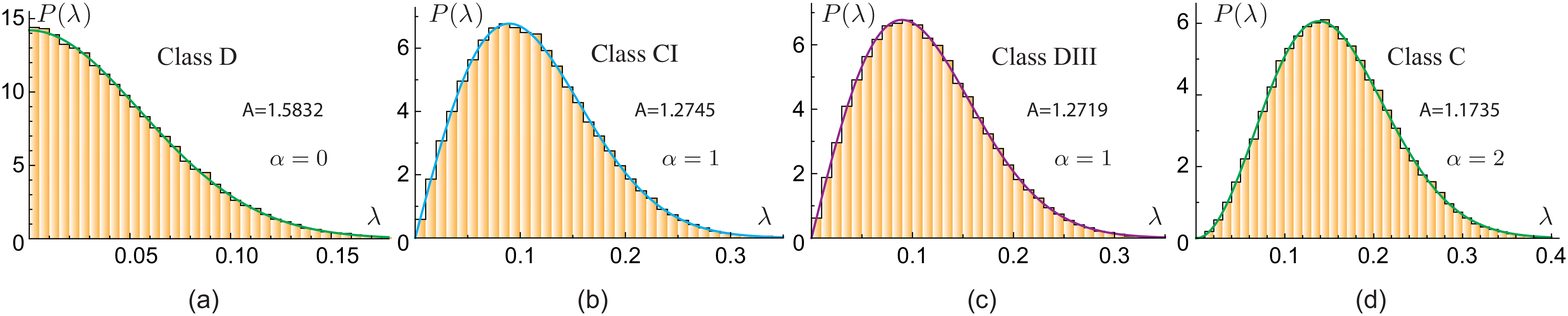}
\caption{The distribution function of the  smallest positive eigenvalue for the 4 BdG Classes
D, CI, DIII and C. The solid line is from the Winger-like surmise obtained in this section.
The Histogram is obtained from the exact diagonalizing of random matrices of size $10^3$
averaged over $10^5$ samples.
}
\label{fig:SM}
\end{figure}

\section{B. Constructing the energy levels by anti-unitary operators in the minimum scheme  }

  In the main text, we counted the energy level degeneracy just from the block structure of the corresponding 10-fold way RMT classes.
  Here, we construct the energy levels systematically and explicitly in terms of the anti-unitary operators,
  so in addition to counting the energy level degeneracy, it also provides the detailed structures of all the energy levels.
  We also discuss the relations among various different basis.

\bigskip
{\bf B1. The $\Lambda$ operator  in the minimum scheme }%---
\bigskip

Here we first summarize the nature of the  operator $ \Lambda= PR $. Because $PR=(-1)^{\lceil N/2\rceil\lfloor N/2\rfloor}RP$,
so it is enough to just focus on $ \Lambda=PR $.

For SYK, $q$ is even, then $q(N + 1)$ is always even:
when $N$ is even, $\Lambda=(-1)^F$ is the conserved parity,
when $N$ is odd, $\Lambda=1$.

For $\mathcal{N}=1$ SYK, $q$ is odd,
when $N$ is even, $q(N+1)$ is odd, $\Lambda=(-1)^F$ is the parity operator
which anti-commutes with $Q$, so is a chirality operator.
When $N$ is odd, $q(N+1)$ is even, $\Lambda=1$.

In short, when $ N $ is even, $ \Lambda $ is the conserved parity operator when $ q $ is even, but a chirality operator
when $ q $ is odd.  When $ N $ is odd, $ \Lambda=1 $ is trivial in any case. Its complete properties is listed in
the table \ref{tab:SYKLambda}.

%Fine tuning the table
\renewcommand{\arraystretch}{1.4}%Default value is 1
\renewcommand\tabcolsep{4pt}%Default value is 0pt

\begin{table}[!htb]
\caption{ The $\Lambda$ operators in the minimum scheme. There are only 3 possibilities: the conserved parity $ (-1)^F $,
the chirality parity $ (-1)^F $ and the trivial identity $ 1 $. }
\begin{tabular}{ c|cccccccc }
\toprule
$N\pmod 8$  &0		&1	&2	    &3	    &4		&5	&6	    &7	\\
\hline
$\Lambda$   &$(-1)^F$	&$1$	&$(-1)^F$   &$1$    &$(-1)^F$	&$1$	&$(-1)^F$   &$1$\\
\hline
$q\pmod2=0$ &$\Lambda$ conserved
	     &trivial
	      &$\Lambda$ conserved
	       &trivial
	        &$\Lambda$ conserved
		 &trivial
		  &$\Lambda$ conserved
		   &trivial	\\
$q\pmod2=1$ &$\Lambda$ chirality
	     &trivial
	      &$\Lambda$ chirality
	       &trivial
	        &$\Lambda$ chirality
		 &trivial
		  &$\Lambda$ chirality
		   &trivial	\\
\hline
Any $q$ &$[P,\Lambda]=0$
	     &$[P,\Lambda]=0$
	      &$\{P,\Lambda\}=0$
	       &$[P,\Lambda]=0$	
	        &$[P,\Lambda]=0$
		 &$[P,\Lambda]=0$
		  &$\{P,\Lambda\}=0$
		   &$[P,\Lambda]=0$\\
Any $q$ &$[R,\Lambda]=0$
	     &$[R,\Lambda]=0$
	      &$\{R,\Lambda\}=0$
	       &$[R,\Lambda]=0$	
	        &$[R,\Lambda]=0$
		 &$[R,\Lambda]=0$
		  &$\{R,\Lambda\}=0$
		   &$[R,\Lambda]=0$\\
\toprule
\end{tabular}
\label{tab:SYKLambda}
\end{table}

\bigskip
{\bf B2. SYK models with $q\pmod 4=0$.}%---
\bigskip

When $ N $ is even, $ \Lambda=PR $ is the conserved parity $ (-1)^F $.
For $N\pmod 8=0,4$, no relation between the two opposite parity blocks.
%thus the ground state degeneracy is equal to internal degeneracy of RMT ensemble;
For $N\pmod 8=2,6$, $P$ swaps between the two opposite parity blocks,
thus there is a one to one correspondence between the energy levels in the two opposite parities.
When $N$ is odd, $ \Lambda=1 $, the Hamiltonian has no block structure.
%thus the ground state degeneracy is determined by the internal degeneracy of RMT ensemble;
In the following, we will determine the energy level degeneracy when $ N $ is even or odd respectively.

{\sl 1. $ N $ is even. }

When $N$ is even, $ \Lambda=PR=(-1)^F $ is the conserved parity.
If choosing $\psi_{+}$ to be an eigenstate with the eigenenergy $E$ and even parity,
because $PR\psi_{+}=\psi_{+}$, $R\psi_{+}\propto P^2R\psi_{+}=P\psi_{+}$,
then one only need clarify the relations between $\psi_{+}$ and $P\psi_{+}$.

When $N\pmod 8=0$, $P$ and $R$ commute with $Q_q$ and preserve parity, $P^2=R^2=+1$.
If $\psi_{+}$ is an eigenstate with eigenenergy $E$ and even parity,
then $P\psi_{+}\propto\psi_{+}$,
thus $Q_q$ has no degeneracy.

When $N\pmod 8=2$, $P$ and $R$ commute with $Q_q$ and swap parity, $P^2=-R^2=+1$.
If $\psi_{+}$ is an eigenstate with the eigenenergy $E$ and even parity,
then $P\psi_{+}$ is an eigenstate with eigenenergy $E$, but odd parity,
thus $Q_q$ has 1+1-fold degeneracy which means the two degenerate states have opposite parities.

When $N\pmod 8=4$, $P$ and $R$ commute with $Q_q$ and preserve parity, and $P^2=R^2=-1$.
If $\psi_{+}$ is an eigenstate with eigenenergy $E$ and even parity,
then $\psi_{+}$ and $P\psi_{+}$
are two orthogonal eigenstate with eigenenergy $E$ and even parity,
thus $Q_q$ has 2-fold degeneracy.

When $N\pmod 8=6$, $P$ and $R$ commute with $Q_q$ and swap parity, and $P^2=-R^2=-1$.
If $\psi_{+}$ is an eigenstate with eigenenergy $E$ and even parity,
then $P\psi_{+}$ is also an eigenstate with eigenenergy $E$, but odd parity,
thus $Q_q$ has 1+1-fold degeneracy.

{\sl 2. $ N $ is odd. }

When $N$ is odd, $ \Lambda=PR=1$,
if choosing $\psi$ to be an eigenstate with eigenenergy $E$,
because $PR\psi=\psi$ and $R\psi\propto P^2R\psi=P\psi$.
Then one still only clarify the relations between $\psi$ and $P\psi$.

When $N\pmod 8=1,7$, $P$ and $R$ commute with $Q_q$, and $P^2=R^2=+1$.
If $\psi$ is an eigenstate with eigenenergy $E$,
then $P\psi\propto\psi$,
thus $Q_q$ has no degeneracy.

When $N\pmod 8=3,5$, $P$ and $R$ commute with $Q_q$, and $P^2=R^2=-1$.
If $\psi$ is an eigenstate with eigenenergy $E$,
then $\psi$ and $P\psi$ are two
orthogonal eigenstates with the same eigenenergy $E$,
thus $Q_q$ has 2-fold degeneracy.

The energy level degeneracy confirms that listed in the 5th row of Table \ref{tab:SYKq}.

%do not remove two blank lines
%\noindent
%Degeneracy with odd $N$ is $1/\sqrt{2}$ of Ref.\cite{You2017}'s,
%is due to different definition of degeneracy for odd case.

\bigskip
{\bf B3. SYK models with $q\pmod 4=2$.}%---
\bigskip

To determine the level degeneracy, we can use the same method as $q\pmod 4=0$ case discussed above.
For any $N$, $P$ and $R$ always anti-commute with $Q_q$.
If $\psi$ is an eigenstate with eigenenergy $E$,
then $P\psi$ is an eigenstate with eigenenergy $-E$,
thus $Q_q$ has no degeneracy, but a spectral mirror symmetry in any case.
In the following, we just spell out the detailed energy level structure.

{\sl 1. $ N $ is even. }

When $N\pmod 8=0$, $P$ and $R$ anti-commute with $Q_q$, preserve parity, and $P^2=R^2=+1$.
If $\psi_{+}$ is an eigenstate with eigenenergy $E$ and even parity,
then $P\psi_{+}$ is an eigenstate with eigenenergy $-E$ and even parity,
thus $Q_q$ has no degeneracy.

When $N\pmod 8=2$, $P$ and $R$ anti-commute with $Q_q$, swap parity, and $P^2=-R^2=+1$.
If $\psi_{+}$ is an eigenstate with eigenenergy $E$ and even parity,
then $P\psi_{+}$ is an eigenstate with eigenenergy $-E$ and odd parity,
thus $Q_q$ has no degeneracy.

When $N\pmod 8=4$, $P$ and $R$ anti-commute with $Q_q$, preserve parity, and $P^2=R^2=-1$.
If $\psi_{+}$ is an eigenstate with eigenenergy $E$ and even parity,
then $P\psi_{+} $ is a eigenstate with  eigenenergy $-E$ and even parity,
thus $Q_q$ has no degeneracy.

When $N\pmod 8=6$,  $P$ and $R$ anti-commute with $Q_q$, swap parity, and $P^2=-R^2=-1$.
If $\psi_{+}$ is an eigenstate with eigenenergy $E$ and even parity,
then $P\psi_{+} $ is an  eigenstate with eigenenergy $-E$ and odd parity,
thus $Q_q$ has no degeneracy.

{\sl 2. $ N $ is odd. }

There is no conserved quantity, $ \Lambda=PR=1 $, $P$, $R$ anti-commute with $Q_q$.
If $\psi$ is an eigenstate with eigenenergy $E$,
then $P\psi=R\psi$ is an eigenstate with eigenenergy $-E$,
thus $Q_q$ has no degeneracy.

The energy level degeneracy confirms that listed in the 7th row of Table \ref{tab:SYKq}.

\bigskip
{\bf B4. SYK supercharge with $q\pmod 4=1$.}%---
\bigskip

In the following, we will discuss the $ N $ even or odd respectively.

{\sl 1. $ N $ is even. }

To determine the energy level degeneracy for $N$ even case,
$  \Lambda=PR= (-1)^F $ is the chirality operator which anti-commutes with $ Q_q $, no conserved quantity exists.
If choosing $\psi$ to be an eigenstate of $Q_q$ with eigenenergy \cite{quasi} $\sqrt{E}$,
then one need clarify the relations among
$\psi$, $P\psi$, $R\psi$, and $PR$.

When $N\pmod 8=0$, %no conserved quantity and
$\{P,Q_q\}=[R,Q_q]=0$ and $P^2=R^2=+1$.
If $\psi$ is an eigenstate of $Q_q$ with eigenenergy $\sqrt{E}$,
( thus $H=Q_q^2$ haa an eigenenergy $E$,)
then $P\psi$ is an eigenstate with eigenenergy $-\sqrt{E}$,
$R\psi\propto\psi$ and $PR\psi\propto P\psi$,
thus $Q_q$ has no degeneracy, but a mirror symmetry,
so $Q_q^2$ has 2-fold degeneracy.

When $N\pmod 8=2$, %no conserved quantity and
$[P,Q_q]=\{R,Q_q\}=0$ and $P^2=-R^2=+1$.
If $\psi$ is an eigenstate of $Q_q$ with eigenenergy $\sqrt{E}$,
then $R\psi$ is an eigenstate with eigenenergy $-\sqrt{E}$,
$P\psi\propto\psi$ and $PR\psi\propto R\psi$,
thus $Q_q$ has no degeneracy, but a mirror symmetry,
so $Q_q^2$ has 2-fold degeneracy.

When $N\pmod 8=4$, %no conserved quantity and
$\{P,Q_q\}=[R,Q_q]=0$ and $P^2=R^2=-1$.
If $\psi$ is an eigenstate of $Q_q$ with eigenenergy $\sqrt{E}$,
then
$\psi$ and $R\psi$ are two orthogonal eigenstates
(namely, $ ( \psi, R\psi )=0  $ )  with eigenenergy $\sqrt{E}$,
$P\psi$ and $PR\psi$ are two orthogonal eigenstates
(namely, $ ( P \psi, PR\psi )=0  $ ) with eigenenergy $-\sqrt{E}$,
thus $Q_q$ has double degeneracy and also a mirror symmetry,
so $Q_q^2$ has 4-fold degeneracy.

When $N\pmod 8=6$, %no conserved quantity and
$[P,Q_q]=\{R,Q_q\}=0$ and $P^2=-R^2=-1$.
If $\psi$ is an eigenstate of $Q_q$ with eigenenergy $\sqrt{E}$,
then
$\psi$ and $P\psi$ are two orthogonal eigenstates with eigenenergy $\sqrt{E}$,
$R\psi$ and $PR\psi$ are two orthogonal eigenstates  with eigenenergy $-\sqrt{E}$,
thus $Q_q$ has double degeneracy and  also a mirror symmetry,
so $Q_q^2$ has 4-fold degeneracy.

{\sl 2. $ N $ is odd. }

To determine level degeneracy for $N$ odd case, since $ \Lambda=PR= 1$,
so $\psi= PR \psi$ and $R\psi\propto P^2 R \psi= P\psi$,
one only need clarify relations between $\psi$ and $P\psi$.

When $N\pmod 8=1$, %$PR\propto I$ and
$[P,Q_q]=[R,Q_q]=0$ and $P^2=R^2=+1$.
In this case,
if $\psi$ is an eigenstate of $Q_q$ with eigenenergy $\sqrt{E}$,
then $P\psi\propto \psi$,
thus $Q_q$ has no degeneracy and no mirror symmetry either,
so $Q_q^2$ has no degeneracy.

When $N\pmod 8=3$, %$PR\propto I$ and
$\{P,Q_q\}=\{R,Q_q\}=0$ and $P^2=R^2=-1$.
If $\psi$ is an eigenstate of $Q_q$ with eigenenergy $\sqrt{E}$,
then $P\psi$ is an eigenstate with eigenenergy $-\sqrt{E}$,
thus $Q_q$ has no degeneracy, but a mirror symmetry,
so $Q_q^2$ has 2-fold degeneracy.

When $N\pmod 8=5$, %$PR\propto I$ and
$[P,Q_q]=[R,Q_q]=0$ and $P^2=R^2=-1$.
If $\psi$ is an eigenstate of $Q_q$ with eigenenergy $\sqrt{E}$,
then $\psi$ and $P\psi$ are two orthogonal eigenstates  with eigenenergy $\sqrt{E}$,
thus $Q_q$ has double degeneracy and no mirror symmetry,
so $Q_q^2$ has 2-fold degeneracy.

When $N\pmod 8=7$, %$PR=I$ and
$\{P,Q_q\}=\{R,Q_q\}=0$ and $P^2=R^2=1$.
If $\psi$ is an eigenstate of $Q_q$ with eigenenergy $\sqrt{E}$,
then $P\psi$ is an eigenstate with eigenenergy $-\sqrt{E}$,
thus $Q_q$ has no degeneracy, but a mirror symmetry,
so $Q_q^2$ has 2-fold degeneracy.

So the energy level degeneracy of $Q_q$ and $ H_q=Q^2_q $ confirm those listed in the 9th and 10th rows of Table \ref{tab:SYKq}.
%However, this counting procedures provide additional results on the detailed structures of the energy levels.
%small inline table

\bigskip
{\bf B5. SYK supercharge with $q\pmod 4=3$.}%---
\bigskip

To determine the level degeneracy, we can use the similar procedure as that in the
above $q\pmod 4=1$ case. So we also discuss $ N $ even and odd respectively.

{\sl 1. $ N $ is even. }

When $N\pmod 8=0$, %no conserved quantity and
$[P,Q_q]=\{R,Q_q\}=0$ and $P^2=R^2=+1$.
If $\psi$ is an eigenstate of $Q_q$ with eigenenergy $\sqrt{E}$,
(thus $H=Q_q^2$ haa an eigenenergy $E$,)
then $R\psi$ is an eigenstate with eigenenergy $-\sqrt{E}$,
$P\psi\propto\psi$ and $PR\psi\propto R\psi$,
thus $Q_q$ has no degeneracy, but a mirror symmetry,
so $Q_q^2$ has 2-fold degeneracy.

When $N\pmod 8=2$, %no conserved quantity and
$\{P,Q_q\}=[R,Q_q]=0$ and $P^2=-R^2=+1$.
If $\psi$ is an eigenstate of $Q_q$ with eigenenergy $\sqrt{E}$,
then $\psi$ and $R\psi$ are two orthogonal eigenstates  with eigenenergy $\sqrt{E}$,
$P\psi$ and $PR\psi$ are two orthogonal eigenstates  with eigenenergy $-\sqrt{E}$,
thus $Q_q$ has double degeneracy and a mirror symmetry,
so $Q_q^2$ has 4-fold degeneracy.

When $N\pmod 8=4$, %no conserved quantity and
$[P,Q_q]=\{R,Q_q\}=0$ and $P^2=R^2=-1$.
if $\psi$ is an eigenstate with eigenenergy $\sqrt{E}$,
then $\psi$ and $P\psi$ are two orthogonal  eigenstates with eigenenergy $\sqrt{E}$,
$R\psi$ and $PR\psi$ are two orthogonal eigenstates with eigenenergy $-\sqrt{E}$,
thus $Q_q$ has double degeneracy and a mirror symmetry,
so $Q_q^2$ has 4-fold degeneracy.

When $N\pmod 8=6$, %no conserved quantity and
$\{P,Q_q\}=[R,Q_q]=0$ and $P^2=-R^2=-1$.
In this case,
if $\psi$ is an eigenstate with eigenenergy $\sqrt{E}$,
then $P\psi$ is an eigenstate with eigenenergy $-\sqrt{E}$, $R\psi\propto\psi$ and $PR\psi\propto P\psi$,
thus $Q_q$ has no degeneracy, but a mirror symmetry,
so $Q_q^2$ has 2-fold degeneracy.

{\sl 2. $ N $ is odd. }

When $N\pmod 8=1$, %$PR\propto I$ and
$\{P,Q_q\}=\{R,Q_q\}=0$ and $P^2=R^2=+1$.
If $\psi$ is an eigenstate with eigenenergy $\sqrt{E}$,
then $P\psi$ is an eigenstate with eigenenergy $-\sqrt{E}$,
thus $Q_q$ has no degeneracy, but a mirror symmetry,
so $Q_q^2$ has 2-fold degeneracy.

When $N\pmod 8=3$, %$PR\propto I$ and
$[P,Q_q]=[R,Q_q]=0$ and $P^2=R^2=-1$.
If $\psi$ is an eigenstate with eigenenergy $\sqrt{E}$,
then $\psi$ and $P\psi$ are two orthogonal eigenstates with eigenenergy $\sqrt{E}$,
thus $Q_q$ has double degeneracy, but no mirror symmetry,
so $Q_q^2$ has 2-fold degeneracy.

When $N\pmod 8=5$, %$PR\propto I$ and
$\{P,Q_q\}=\{R,Q_q\}=0$ and $P^2=R^2=-1$.
If $\psi$ is an eigenstate with eigenenergy $\sqrt{E}$,
then $P\psi$ is an eigenstate with eigenenergy $-\sqrt{E}$,
thus $Q_q$ has no degeneracy, but a mirror symmetry,
so $Q_q^2$ has 2-fold degeneracy.

When $N\pmod 8=7$, %$PR\propto I$ and
$[P,Q_q]=[R,Q_q]=0$ and $P^2=R^2=+1$.
If $\psi$ is a eigenstate with eigenenergy $\sqrt{E}$,
then $P\psi\propto\psi$,
thus $Q_q$ has no degeneracy and no mirror symmetry either,
so $Q_q^2$ has no degeneracy.

So the energy level degeneracy of $Q_q$ and $ H_q=Q^2_q $ confirm those listed in the 12nd and 13rd rows of Table \ref{tab:SYKq}.
%However, this counting procedures provide additional results on the detailed structures of the energy levels.	

\section{C. Classifying SYK models with odd $N$ in enlarged Hilbert space by adding a Majorana fermion: odd $ N $ case }

%{\color{blue}\textit{Relation to other scheme.}}---

%

 When $N$ is odd, following \cite{You2017,hybrid}, one may
add a Majorana fermion $\chi_\infty=\chi_{N+1}$ to the system.
Then the Hilbert space has dimension $2^{N_c}$
which is twice that of the minimum scheme used in the main text.
In fact, as shown below, there are two more conserved quantities which all commute with the Hamiltonian, but not commute with each other.
So there is only one more compatible conserved quantity.

%Following the convention introduced by

 In the minimal scheme used in the main text, one must choose $ \chi_{2i-1} $ to be real,
 $ \chi_{2i} $ to be imaginary \cite{footnote1}. In the enlarged Hilbert space, one can choose either way.
 To be contrasted and also complementary to that used in the main text, we choose $ \chi_{2i} $ to be real,
 while $ \chi_{2i-1} $ to be imaginary. The two different choices lead to quite different classification schemes, but to
 the same results after interpreting  the results carefully. In this section, especially Sec.C2-1,
 we will establish the main differences and also intricate connections  between  the two choices.
 Then the $N_c=(N+1)/2$ complex fermion creation and annihilation operators $c_i^\dagger$ and $c_i$
 are defined as:
\begin{align}
	c_i=(\chi_{2i}-i\chi_{2i-1})/2,\quad
	c_i^\dagger=(\chi_{2i}+i\chi_{2i-1})/2,
	\qquad i=1,\cdots N_c=(N+1)/2
\label{IR}
\end{align}

Then the fermion number operator $F=\sum_{i=1}^{N_c}c_i^\dagger c_i$.
One can  define two particle-hole symmetry operators to be \cite{Fu2016,You2017,Cotler2017,hybrid}
\begin{align}
    P=K\prod_{i=1}^{N_c}(c_i^\dagger+c_i),\quad
    R=K\prod_{i=1}^{N_c}(c_i^\dagger-c_i)
\label{eq:CPR}
\end{align}

%These complex fermions can be represented as real matrices
%by adopting the Jordan-Wigner construction[cite JHEP09(2017)050],
%so that $Kc_iK=c_i$ and $Kc_i^\dagger K=c_i^\dagger$.

There is also another useful anti-unitary operator $Z$
by factoring out the extra Majorana fermion
from the particle-hole symmetry operator $P$,
which can be written as \cite{You2017,hybrid}
\begin{align}
    P=Z\chi_{\infty},\qquad
    Z=P\chi_{\infty}=K\prod_{i=1}^{N_c-1}(c_i^\dagger+c_i)\>.
\end{align}

It can be shown that \cite{hybrid,color}
\begin{alignat*}{3}
    &P\chi_i P^{-1}=(-1)^{N_c-1}\chi_i,\qquad
	&&R\chi_i R^{-1}=(-1)^{N_c}\chi_i,\quad
	    &&Z\chi_i Z^{-1}=(-1)^{N_c}\chi_i,\quad
		i=1,\cdots,N\\
    &P\chi_\infty P^{-1}=(-1)^{N_c-1}\chi_\infty,\qquad
	&&R\chi_\infty R^{-1}=(-1)^{N_c}\chi_\infty,\quad
	    &&Z\chi_\infty Z^{-1}=(-1)^{N_c-1}\chi_\infty,
\end{alignat*}
 which leads to:
\begin{align}
    PQ_qP^{-1}=(-1)^{\lfloor q/2\rfloor}(-1)^{q(N_c-1)}Q_q,\quad
    RQ_qR^{-1}=(-1)^{\lfloor q/2\rfloor}(-1)^{qN_c}Q_q,\quad
    ZQ_qZ^{-1}=(-1)^{\lfloor q/2\rfloor}(-1)^{qN_c}Q_q,
\end{align}
 One may also find their squared values
\begin{align}
    P^2=(-1)^{\lfloor N_c/2\rfloor},\quad
    R^2=(-1)^{\lfloor 3N_c/2\rfloor},\quad
    Z^2=(-1)^{\lfloor (N_c-1)/2\rfloor}.
\label{squared}
\end{align}

In fact, one may also construct other anti-unitary operators such as adding $\chi_\infty$ to $R$,
which leads to another  anti-unitary operator $
    Z'=R\chi_{\infty}
	=K\Big[\prod_{i=1}^{N_c}(c_i^\dagger-c_i)\Big]\chi_\infty\> $.
However, $Z'$ can be expressed by a product of $R$, $P$, and $Z$,
so $R$, $P$, and $Z$ will be the complete set of anti-unitary operators to perform the classifications.
The relation between the anti-unitary operators and RMT classes is listed in Table S1.

%Fine tuning the table
\renewcommand{\arraystretch}{1.4}%Default value is 1
\renewcommand\tabcolsep{6pt}%Default value is 0pt

\begin{table}[!htb]
\caption{The energy level degeneracy of SYKq in the extended scheme at odd $ N $.
 When $ N $ is even, there is no extended scheme, so we only list odd $ N $ case here.
 When $ q $ is even, $ 1+1 $ means the two energy levels
at the two opposite parity $ (-1)^F $ sector. When $ q $ is odd,  for the supercharge $ Q_q $,
it is at the two opposite parity $ (-1)^F i \chi_{\infty} $ sector. There is always a mirror symmetry,
so the degeneracy of $ H_q $ is always twice of $ Q_q  $. How to compare this table with Table \ref{tab:SYKq}
in the main text is discussed in Sec. C6. }
\begin{tabular}{ c|cccccccc }
\toprule
$N\pmod 8$				    &1	&3	&5	&7	  \\
\hline
$q\pmod4=0,~~ Q_q $		&1+1	&2+2	&2+2	&1+1	\\
$q\pmod4=2,~~ Q_q $	 &1+1	&1+1	& 1+1	&1+1		\\
\hline
$q\pmod4=1,~~ Q_q $	&1	& 1+1	& 2	&  1+1		\\
$q\pmod4=1,~~ H_q $	&2	&4	& 4	&4		\\
$q\pmod4=3,~~ Q_q $	&1+1	&2	&1+1	& 1		\\
$q\pmod4=3,~~ H_q $	&4	&4	&4	&2		\\
\toprule
\end{tabular}
\label{tab:ext}
\end{table}

\bigskip
{\bf C1. The systematic approach in the enlarged Hilbert space. }
\bigskip

Because $ N $ is fixed to be odd in this section, so in the following, we will discuss $ q $ even and odd respectively.

{\sl 1. $ q $ is even }

The total fermion number parity $(-1)^F$ is a conserved quantity.
The commutation relations between the anti-unitary operators and the conserved parity
can be easily worked out
\begin{align}
	P(-1)^FP^{-1}=(-1)^{N_c}(-1)^F,\quad
	R(-1)^FR^{-1}=(-1)^{N_c}(-1)^F,\quad
	Z(-1)^FZ^{-1}=(-1)^{N_c-1}(-1)^F
\label{eqSM:PRZF}
\end{align}

  From the expressions,
\begin{align}
    PR=i^{N_c}\chi_2\chi_{4}\cdots\chi_{2N_c}\chi_1\chi_{3}\cdots\chi_{2N_c-1},
	\quad
    (-1)^F=(i\chi_1\chi_2)(i\chi_3\chi_4)\cdots(i\chi_{2N_c-1}\chi_{2N_c})
\end{align}
  one can see $(-1)^F=(-1)^{N_c(N_c+1)/2}PR=(-1)^{\lceil N_c/2\rceil}PR$
  which can be simply written as $(-1)^F\propto PR$.

When doing the classification, one need use the anti-unitary operators which preserves the parity $(-1)^F$,
then obtain the classification from their  commutation relations with $Q_q$, also their squared values by comparing with Table. S1.
When counting the  energy level degeneracy, if $\psi_{+}$ is an eigenstate of $Q_q$ with the eigenenergy $E$ and even parity,
then the degeneracy can be obtained by finding the
relations among $\psi_{+}$, $P\psi_{+}$, $Z\psi_{+}$, and $ZP\psi_{+}$.
Because $(-1)^F=(-1)^{\lceil N_c/2\rceil}PR$, it is easy to show that
$PR\psi_{+}\propto \psi_{+}$, $R\psi_{+}\propto P\psi_{+}$,
$ZR\psi_{+}\propto ZP\psi_{+}$, and $ZPR\psi_{+}\propto Z\psi_{+}$.
Due to $PR=\pm RP$ and $ZPR=\pm PZR$, different orderings of $P, R, Z$ in these states
will not lead to any new states.

{\sl 2.  $ q $ is odd }

The total fermion number parity $(-1)^F$ and $\chi_{\infty}$
anti-commute with $Q_q$,
thus $(-1)^Fi\chi_{\infty}$ commutes with $Q_q$.
Notice, here $i$ is a necessary factor to make $(-1)^Fi\chi_{\infty}$ Hermitian.
Since $[(-1)^Fi\chi_{\infty}]^2=(-1)^{2F}=1$, we will still call it ``parity''.
One can check its relation with $P$, $R$ and $Z$
\begin{align}
	P(-1)^Fi\chi_{\infty}P^{-1}=+(-1)^F\chi_{\infty},\quad
	R(-1)^Fi\chi_{\infty}R^{-1}=-(-1)^F\chi_{\infty},\quad
	Z(-1)^Fi\chi_{\infty}Z^{-1}=-(-1)^F\chi_{\infty}
\label{eqSM:PRZFi}
\end{align}

For the classification, one need use the anti-unitary operators which commute with
$(-1)^Fi\chi_{\infty}$,
 then obtain the classification by checking their commutation relations with $Q_q$, their squared value by comparing with Table. S1.
For the energy level degeneracy,
if $\psi_{+}$ is an eigenstate of $Q_q$ with eigenenergy $E$ and even ``parity'',
then the degeneracy can be obtained by clarifying
the relations among $\psi_{+}$, $P\psi_{+}$, $Z\psi_{+}$, and $ZP\psi_{+}$.
Note that due to the fact $(-1)^Fi\chi_{\infty}=(-1)^{\lfloor N_c/2\rfloor}iZR$, one can see
that  $ZR\psi_{+}\propto \psi_{+}$, $R\psi_{+}\propto Z\psi_{+}$,
$PR\psi_{+}\propto ZP\psi_{+}$, and $ZPR\psi_{+}\propto P\psi_{+}$.

In the following, we are ready to discuss the classification and level degeneracy.

\bigskip
{\bf C2. $q\pmod4=0$ and $ N $ odd   }
\bigskip

Then $ P, R, Z $ all commute with $Q_q$, $[P,Q_q]=[R,Q_q]=[Z,Q_q]=0$, thus play the role as $T_+$.

When $N\pmod 8=1$, $N_c\pmod 4=1$,
$P$ and $R$ swap parity, $Z$ preserves parity, $P^2=-R^2=Z^2=+1$, one must use $ Z $ to do the classification,
thus $Q_q$ belongs to Class AI(GOE).
If $\psi_{+}$ is an eigenstate of $Q_q$
with an eigenenergy $E$ and even parity,
then $\psi_{+}$ and $Z\psi_{+}$
are the same eigenstate with eigenenergy $E$ and even parity,
while $P\psi_{+}$ and $ZP\psi_{+}$
are the same eigenstate with eigenenergy $E$ and odd parity,
thus $Q_q$ has 1+1-fold degeneracy.

When $N\pmod 8=3$, $N_c\pmod 4=2$,
$P$ and $R$ preserve parity, $Z$ swaps parity, $P^2=R^2=-Z^2=-1$,
one can use either $ P $ or $ R $ to do the classification,
thus $Q_q$ belongs to Class AII(GSE).
If $\psi_{+}$ is an eigenstate of $Q_q$
with eigenenergy $E$ and even parity,
then $\psi_{+}$ and $P\psi_{+}$
are two orthogonal eigenstates with eigenenergy $E$ and even parity,
$Z\psi_{+}$ and $ZP\psi_{+}$
are two orthogonal eigenstates with eigenenergy $E$ and odd parity,
thus $Q_q$ has 2+2-fold degeneracy.

When $N\pmod 8=5$, $N_c\pmod 4=3$,
$P$ and $R$ swap parity, $Z$ preserves parity, $P^2=-R^2=Z^2=-1$,
one must use $ Z $ to do the classification,
thus $Q_q$ belongs to Class AII(GSE).
If $\psi_{+}$ is an eigenstate of $Q_q$
with eigenenergy $E$ and even parity,
then $\psi_{+}$ and $Z\psi_{+}$
are two orthogonal eigenstates with eigenenergy $E$ and even parity,
$P\psi_{+}$ and $ZP\psi_{+}$
are two orthogonal eigenstates with eigenenergy $E$ and odd parity,
thus $Q_q$ has 2+2-fold degeneracy.

When $N\pmod 8=7$, $N_c\pmod 4=0$,
$P$ and $R$ preserve parity, $Z$ swaps parity, $P^2=R^2=-Z^2=+1$,
one can use either $ P $ or $ R $ to do the classification,
thus $Q_q$ belongs to Class AI(GOE).
If $\psi_{+}$ is an eigenstate of $Q_q$
with an eigenenergy $E$ and even parity,
then $\psi_{+}$ and $P\psi_{+}$ are the same eigenstate
with eigenenergy $E$ and even parity,
$Z\psi_{+}$ and $ZP\psi_{+}$ are the same eigenstate
with eigenenergy $E$ and odd parity,
thus $Q_q$ has 1+1-fold degeneracy.

These results are listed in the 2nd row in Table \ref{tab:ext}.
Here, we perform the classification with the conserved parity $ (-1)^F $ which is also Hermitian.
In fact, there is another conserved quantity $ \chi_{\infty} $ which is also Hermitian.
So one can also do the classification by using  $ \chi_{\infty} $ and indeed reach the same conclusions.
Note that choosing different conserved quantity, one
may need to use different anti-unitary symmetry operators to do classification.
However, because $ \{ \chi_{\infty},  (-1)^F \} =0 $, so they do not commute. One can only use one of the two or
their combination to do the classification as shown in the following.

%When $N_c$ even, $N\pmod 8=3,7$, $P$ and $R$ preserve parity
%and $[P,Q_q]=[R,Q_q]=0$ and $P^2=R^2=(-1)^{N_c/2}=-1,1$,
%thus $Q_q$ belongs to Class AII,AI, respectively.

%When $N_c$ odd, $N\pmod 8=1,5$, $Z$ preserve parity
%and $[Z,Q_q]=0$ and $Z^2=(-1)^{(N_c-1)/2}=+1,-1$,
%thus $Q_q$ belongs to Class AI, AII, respectively.

\bigskip
\noindent
{\sl 1. The impossibility to do the classifications without touching $ \chi_{\infty} $ in the basis \ref{IR}. }

One may would like to just use $R$ and $Z$ to classify
which do not involve $\chi_\infty$.
However, in the basis defined by Eq.\ref{IR},
the classification can not be done without touching $\chi_\infty$ as shown in the following.

%Since we have $(N+1)/2$ pure imaginary matrix and $(N-1)/2$ real matrix,
%which suggest Hilbert space dimension need to be $2^{N_c}=2^{(N+1)/2}$,
As shown above, there are two mutually anti-commuting conserved quantities $ \chi_{\infty},  (-1)^F $, one can
form another conserved quantity $iZR=(-1)^{\lfloor N_c/2\rfloor}(-1)^Fi\chi_\infty$ which is just the product of the two.
But $ \chi_{\infty} $ gets canceled out in the product. So it is tempting to see if one can do the classification using
the anti-unitary operators $ R, Z $ and this conserved quantity, all of which do not involve $\chi_\infty$.

When $N$ is odd, we have $[Z,R]=0$
which lead to the commutation relation between $Z$, $R$, $P$ and the conserved quantity $iZR$
\begin{align}
	ZiZRZ^{-1}=-iZR,\quad
	RiZRR^{-1}=-iZR,\quad
	PiZRP^{-1}=iZR
\end{align}

Thus $Z$ and $R$ can not be used to classify.
One must use $P$, which preserves $iZR$, commute with Hamiltonian
and $P^2=(-1)^{\lfloor N_c/2\rfloor}=(-1)^{\lfloor (N+1)/4\rfloor}$.
So we reach the same classification. This is expected as stated in \cite{footnote1}, so
$ \chi_{\infty} $ is unavoidable in the basis \cite{IR}.

However, if one redefine Eq.\ref{IR} as $c_i=(\chi_{2i-1}-i\chi_{2i})/\sqrt{2}$ by choosing $\chi_{2i-1}$ real and $\chi_{2i}$ imaginary,
one may just use $R$ and $Z$ to classify which do not touch $\chi_\infty$.  Of course, this is expected, because that is the basis \cite{footnote1} we used in the main text where one can perform the classification in the minimal Hilbert space  without adding
the Majorana fermions at $ \infty $. It was designed not to touch ( in fact, no need to add ) $ \chi_{\infty} $ at first place.

\bigskip
{\bf C3.  $q\pmod4=2$ and odd $ N $. }
\bigskip

Then $ P, R, Z $ all anti-commute with $Q_q$, $\{P,Q_q\}=\{R,Q_q\}=\{Z,Q_q\}=0$, thus play role as $T_-$.

%When $N_c$ even, $N\pmod 8=3,7$, $P$ and $R$ preserve parity
%and $\{P,Q_q\}=\{R,Q_q\}=0$ and $P^2=R^2=(-1)^{N_c/2}=-1,1$,
%thus $Q_q$ belongs to Class C,D, respectively.

%When $N_c$ odd, $N\pmod 8=1,5$, $Z$ preserve parity
%and $\{Z,Q_q\}=0$ and $Z^2=(-1)^{\lfloor (N_c-1)/2\rfloor}=+1,-1$,
%thus $Q_q$ belongs to Class D,C, respectively.

When $N\pmod 8=1$, $N_c\pmod 4=1$, $P$ and $R$ swap parity, $Z$ preserves parity, $P^2=-R^2=Z^2=+1$.
So one must use $ Z $ to do the classification,
thus $Q_q$ belongs to Class D(BdG).
If $\psi_{+}$ is an eigenstate of $Q_q$
with the eigenenergy $E$ and  even parity,
then $\psi_{+}$ and $ZP\psi_{+}$
are two orthogonal eigenstates with the same eigenenergy $E$,
but opposite parities,
$P\psi_{+}$ and $Z\psi_{+}$
are two orthogonal eigenstate with the same eigenenergy $-E$,
but opposite parities,
thus $Q_q$ has 1+1-fold degeneracy and also a mirror symmetry.

When $N\pmod 8=3$, $N_c\pmod 4=2$,
$P$ and $R$ preserve parity, $Z$ swaps parity, $P^2=R^2=-Z^2=-1$,
So one can use either $ P $ or $ R $ to do the classification,
thus $Q_q$ belongs to Class C(BdG).
The relations among $\psi_{+}$, $P\psi_{+}$, $Z\psi_{+}$, and $ZP\psi_{+}$
are the same as those in the $N\pmod 8=1$ case,
thus $Q_q$ has 1+1-fold degeneracy and also a mirror symmetry.

When $N\pmod 8=5$, $N_c\pmod 4=3$,
$P$ and $R$ swap parity, $Z$ preserves parity, and $P^2=R^2=Z^2=-1$.
So one must use $ Z $ to do the classification,
thus $Q_q$ belongs to Class C(BdG).
The relations among $\psi_{+}$, $P\psi_{+}$, $Z\psi_{+}$, and $ZP\psi_{+}$
are the same as those in the $N\pmod 8=1$ case,
thus $Q_q$ has 1+1-fold degeneracy and a mirror symmetry.

When $N\pmod 8=7$, $N_c\pmod 4=0$,
$P$ and $R$ preserve parity, $Z$ swap parity, $P^2=R^2=-Z^2=+1$.
So one can use either $ P $ or $ R $ to do the classification,
thus $Q_q$ belongs to Class D(BdG).
The relations among $\psi_{+}$, $P\psi_{+}$, $Z\psi_{+}$, and $ZP\psi_{+}$
are the same as those in the $N\pmod 8=1$ case,
thus $Q_q$ has 1+1-fold degeneracy and a mirror symmetry.

These results are listed in the 3rd row in Table \ref{tab:ext}.

%When $N_c$ even, $N\pmod 8=3,7$, only $P$ preserve parity
%and $\{P,Q_q\}=0$ and $P^2=(-1)^{\lfloor N_c/2\rfloor}=-1,+1$,
%thus $Q_q$ belongs to Class C, D, respectively.

%When $N_c$ odd, $N\pmod 8=1,5$, only $P$ preserve parity
%and $[P,Q_q]=0$ and $P^2=(-1)^{\lfloor N_c/2\rfloor}=+1,-1$,
%thus $Q_q$ belongs to Class AI, AII, respectively.

\bigskip
{\bf C4. $q\pmod4=1$ and odd $ N $ }
\bigskip

When $N\pmod 8=1$, $N_c\pmod 4=1$,
$P$ preserves parity, $R$ and $Z$ swap parity,
$[P,Q_q]=\{R,Q_q\}=\{Z,Q_q\}=0$, and $P^2=-R^2=Z^2=+1$,
thus $Q_q$ belongs to Class AI(GOE).
If $\psi_{+}$ is an eigenstate of $Q_q$ with eigenenergy $\sqrt{E}$ and even ``parity'',
then $\psi_{+}$ and $P\psi_{+}$
are the same eigenstate with an eigenenergy $\sqrt{E}$ and even ``parity'',
$ZP\psi_{+}$ and $Z\psi_{+}$
are the same eigenstate with eigenenergy $-\sqrt{E}$ and odd ``parity'',
thus $Q_q$ has no degeneracy, but a mirror symmetry,
so $Q_q^2$ has 2-fold degeneracy.

When $N\pmod 8=3$, $N_c\pmod 4=2$,
$P$ preserves parity, $R$ and $Z$ swap parity,
$\{P,Q_q\}=[R,Q_q]=[Z,Q_q]=0$, and $P^2=R^2=-Z^2=-1$,
thus $Q_q$ belongs to Class C(BdG).
If $\psi_{+}$ is an eigenstate of $Q_q$
with eigenenergy $\sqrt{E}$ and even ``parity'',
then $\psi_{+}$ and $Z\psi_{+}$
are two orthogonal eigenstates with eigenenergy $\sqrt{E}$, but opposite ``parity'',
$ZP\psi_{+}$ and $P\psi_{+}$
are two orthogonal eigenstates with eigenenergy $-\sqrt{E}$, but opposite ``parity'',
thus $Q_q$ has 1+1 degeneracy and also a mirror symmetry,
so $Q_q^2$ has 4-fold degeneracy.

When $N\pmod 8=5$, $N_c\pmod 4=3$,
$P$ preserves parity, $R$ and $Z$ swap parity,
$[P,Q_q]=\{R,Q_q\}=\{Z,Q_q\}=0$, and $P^2=-R^2=Z^2=-1$,
thus $Q_q$ belongs to Class AII(GSE).
If $\psi_{+}$ is an eigenstate of $Q_q$
with eigenenergy $\sqrt{E}$ and even ``parity'',
then
$\psi_{+}$ and $P\psi_{+}$
are two orthogonal eigenstates with eigenenergy $\sqrt{E}$ and even ``parity'',
$ZP\psi_{+}$ and $Z\psi_{+}$
are two orthogonal eigenstates with eigenenergy $-\sqrt{E}$ and odd ``parity'',
thus $Q_q$ has double degeneracy and also a mirror symmetry,
so $Q_q^2$ has 4-fold degeneracy.

When $N\pmod 8=7$, $N_c\pmod 4=0$,
$P$ preserves parity, $R$ and $Z$ swap parity,
$\{P,Q_q\}=[R,Q_q]=[Z,Q_q]=0$, and $P^2=R^2=-Z^2=+1$,
thus $Q_q$ belongs to Class D(BdG).
If $\psi_{+}$ is an eigenstate of $Q_q$
with and eigenenergy $\sqrt{E}$ and  even ``parity'',
then $\psi_{+}$ and $Z\psi_{+}$
are two orthogonal eigenstates with eigenenergy $\sqrt{E}$, but opposite ``parity'',
$ZP\psi_{+}$ and $P\psi_{+}$
are two orthogonal eigenstates with eigenenergy $-\sqrt{E}$, but opposite ``parity'',
thus $Q_q$ has 1+1 degeneracy and a mirror symmetry,
so $Q_q^2$ has 4-fold degeneracy.

These results for $ Q_q $ and $ H_q= Q^2_q $ are listed in the 4th and 5th rows in Table \ref{tab:ext} respectively.

\bigskip
{\bf C5. $q\pmod4=3$ and odd $ N $. }
\bigskip

%When $q\pmod4=3$ and $N$ is odd,

The situation is similar to $q\pmod4=1$ case.

%When $N_c$ even, $N\pmod 8=3,7$, only $P$ preserve parity
%and $[P,Q_q]=0$ and $P^2=(-1)^{\lfloor N_c/2\rfloor}=-1,+1$,
%thus $Q_q$ belongs to Class AII, AI, respectively.

%When $N_c$ odd, $N\pmod 8=1,5$, only $P$ preserve parity
%and $\{P,Q_q\}=0$ and $P^2=(-1)^{\lfloor N_c/2\rfloor}=+1,-1$,
%Thus $Q_q$ belongs to Class D,C, respectively.

When $N\pmod 8=1$, $N_c\pmod 4=1$,
$P$ preserves parity, $R$ and $Z$ swap parity,
$\{P,Q_q\}=[R,Q_q]=[Z,Q_q]=0$, and $P^2=-R^2=Z^2=+1$,
thus $Q_q$ belongs to Class D(BdG).
If $\psi_{+}$ is an eigenstate of $Q_q$
with an eigenenergy $\sqrt{E}$ and even ``parity'',
then $\psi_{+}$ and $Z\psi_{+}$
are two orthogonal eigenstates with eigenenergy $+\sqrt{E}$, but opposite ``parity'',
$P\psi_{+}$ and $ZP\psi_{+}$
are two orthogonal  eigenstates with eigenenergy $-\sqrt{E}$, but opposite ``parity'',
thus $Q_q$ has 1+1 degeneracy and also a mirror symmetry,
so $Q_q^2$ has 4-fold degeneracy.

When $N\pmod 8=3$, $N_c\pmod 4=2$,
$P$ preserves parity, $R$ and $Z$ swap parity,
$[P,Q_q]=\{R,Q_q\}=\{Z,Q_q\}=0$, and $P^2=R^2=-Z^2=-1$,
thus $Q_q$ belongs to Class AII(GSE).
If $\psi_{+}$ is an eigenstate of $Q_q$
with eigenenergy $\sqrt{E}$ and even ``parity'',
then $\psi_{+}$ and $P\psi_{+}$
are two orthogonal eigenstates with eigenenergy $+\sqrt{E}$ and even ``parity'',
$Z\psi_{+}$ and $ZP\psi_{+}$
are two orthogonal eigenstates with eigenenergy $-\sqrt{E}$ and odd ``parity'',
thus $Q_q$ has double degeneracy and also a mirror symmetry,
so $Q_q^2$ has 4-fold degeneracy.

When $N\pmod 8=5$, $N_c\pmod 4=3$,
$P$ preserves parity, $R$ and $Z$ swap parity,
$\{P,Q_q\}=[R,Q_q]=[Z,Q_q]=0$, and $P^2=-R^2=Z^2=-1$,
thus $Q_q$ belongs to Class C(BdG).
If $\psi_{+}$ is an eigenstate of $Q_q$
with eigenenergy $\sqrt{E}$ and even ``parity'',
then $\psi_{+}$ and $Z\psi_{+}$
are two orthogonal eigenstates with eigenenergy $+\sqrt{E}$, but opposite ``parity'',
$ZP\psi_{+}$ and $P\psi_{+}$
are two orthogonal eigenstates with eigenenergy $-\sqrt{E}$, but opposite ``parity'',
thus $Q_q$ has 1+1 degeneracy and also a mirror symmetry,
so $Q_q^2$ has 4-fold degeneracy.

When $N\pmod 8=7$, $N_c\pmod 4=0$,
$P$ preserves parity, $R$ and $Z$ swap parity,
$[P,Q_q]=\{R,Q_q\}=\{Z,Q_q\}=0$, and $P^2=R^2=-Z^2=+1$,
thus $Q_q$ belongs to Class AI(GOE).
If $\psi_{+}$ is an eigenstate of $Q_q$
with eigenenergy $\sqrt{E}$ and  even ``parity'',
then $\psi_{+}$ and $P\psi_{+}$
are the same eigenstate with eigenenergy $+\sqrt{E}$ and even ``parity'',
$Z\psi_{+}$ and $ZP\psi_{+}$
are the same eigenstates with eigenenergy $-\sqrt{E}$ and odd``parity'',
thus $Q_q$ has no degeneracy, but a mirror symmetry,
so $Q_q^2$ has 2-fold degeneracy.

These results for $ Q_q $ and $ H_q= Q^2_q $ are listed in the 6th and 7th rows in Table \ref{tab:ext} respectively.

\bigskip
{\bf C6. Comparison of Table \ref{tab:ext} of the extended scheme at odd $ N $  to Table.\ref{tab:SYKq} of the minimum scheme in the main text }
\bigskip

When the total number of Majorana fermion $ N $ is even,
it is straightforward and convenient to pair two Majorana fermions to form a Dirac fermion
\cite{Fidkowski2011,You2017,Cotler2017,Fu2016}.
This construction has the same Hilbert space dimension $ 2^{N/2} $ and also same sets of symmetry operators as that with the Majorana fermions
in the minimum scheme used in the main text, thus the two schemes are equivalent.
This is why we only list odd $ N $ in Fig.\ref{tab:ext}.

However, when the total number of Majorana fermion $ N $ is odd,
As shown in this section, it is typical and intuitive to define a enlarged Hilbert space by adding an extra Majorana fermion
that is decoupled from the ones appearing in the Hamiltonian
\cite{Fidkowski2011,You2017,hybrid},
thus the dimension of an enlarged Hilbert space is twice that used in the main text.
The relation between the two can be understood from
the $2^{(N+1)/2}$-dimensional representation of Majoranas,
then the $Q_q$ is automatically in a block diagonal form  for even $q$
or block off-diagonal form for odd $q$, each block is identical to our $2^{(N-1)/2}$-dimensional
minimal representation of $Q_q$,
thus doubled the level degeneracy of $Q_q$ (for even $q$) and $Q_q^2$ (for odd $q$).
In the following, we will show that this is indeed the case when comparing
Table \ref{tab:SYKq} in the minimum scheme to Table \ref{tab:ext} in the extended scheme.

{\sl (1)  even $ q $ }

When $q\pmod4=0 $, the level degeneracy $ d=1,2,2,1 $ in the minimum scheme at $ N=1,3,5, 7 $ respectively in Table \ref{tab:ext} which
is identical to  $ d=1,2,2,1 $ in the same parity sector in extended scheme in Table \ref{tab:SYKq}.

When $q\pmod4=2 $, the level degeneracy $ d=1,1,1,1 $ in the minimum scheme at $ N=1,3,5, 7 $ respectively in Table \ref{tab:ext} which
is identical to  $ d=1,1,1,1 $ in the same parity sector in extended scheme in Table \ref{tab:SYKq}.

As a result, the degeneracy of $Q_q$ in the enlarged Hilbert space is twice that
in the minimum Hilbert space for even $q$.

{\sl (2)  odd $ q $ }

For odd $ q $, there is always an extra partner of eigenenergy with negative energy.
If $Q_q$ with odd $q$ has no mirror symmetry in the minimum scheme,
then adding an extra Majorana fermions makes $Q_q$ to have a spectral mirror symmetry in enlarged Hilbert space.
Then the degeneracy of $ Q_q $ is the same in both schemes,
but the degeneracy of $Q_q^2$ will be doubled in enlarged Hilbert space.
If $Q_q$ with odd $q$ has a mirror symmetry already in the minimum scheme,
then the degeneracy of $ Q_q $ in the minimum scheme is the same as that projected to the same parity $(-1)^Fi\chi_{\infty}$ sector.
The degeneracy of $Q_q^2$ will also be doubled in enlarged Hilbert space.

When $q\pmod4=1 $, the level degeneracy  $ d=1,1,2,1 $ of $ Q_q $  with ( no,yes,no,yes) mirror symmetry at $ N=1,3,5, 7 $
in the minimum scheme  respectively
is identical to  $ d=1,2,2,1 $ of $ Q_q $  with ( no,yes,no,yes) projection  in the extended scheme.
Then the degeneracy of $ H_q=Q^2_q $ is $ d_H=1,2,2,2 $ in the minimum scheme which is twice enlarged to
$ d_H=2,4,4,4 $ in the extended scheme.

When $q\pmod4=3 $, the level degeneracy  $ d=1,2,1,1 $ of $ Q_q $  with ( yes,no,yes, no ) mirror symmetry at $ N=1,3,5, 7 $
in the minimum scheme  respectively
is identical to  $ d=1,2,1,1 $ of $ Q_q $  with ( yes,no,yes,no) projection  in the extended scheme.
Then the degeneracy of $ H_q=Q^2_q $ is $ d_H=2,2,2,1 $ in the minimum scheme which is twice enlarged to
$ d_H=4,4,4,2 $ in the extended scheme.

%It explains why our degeneracy is always one half of other's, when $N$ is odd.
%For level degeneracy per boundary or quantum dimension (qdim) defined in \cite{You2017},
%which is defined to be the square root of the level degeneracy with both boundaries considered,
%thus relation between different definitions becomes $\text{qdim}=\sqrt{2}d_\text{our}=d_\text{enlarged}/\sqrt{2}$.

As a comparison, we also perform exact diagonalization in the enlarged Hilbert
which lead to the same results presented in Fig.1-3 in the main text.

%\clearpage

\section{D. Counting the degeneracy of the $\mathcal{N}=1$ supersymmetric SYK model in $ (-1)^F $ basis: Odd $ q $ case }

\begin{table}[!htb]
\caption{ The energy level degeneracy of $\mathcal{N}=1$ supersymmetric SYK $ H_q $ in $ (-1)^F $ basis.
 When $ N $ is even, $ 1+1 $ means in the two opposite parities of $ (-1)^F $.
 When $ N $ is odd, $ 1+1=1_{+,+}+1_{-,-}, 1+1+1+1=1_{+,+}+1_{+,-}+ 1_{-,+}+1_{-,-}, 2+2=2_{+,+}+2_{-,-} $.
 The first of $ (\pm,\pm) $ is $ (-1)^F $, the second is $ i\chi_\infty Q_q $.
 It is compared to the previous two tables \ref{tab:SYKq} and  \ref{tab:ext} in Sec. D6. }
\begin{tabular}{ c|c|c|c|c|c|c|c|c }
\toprule
$N\pmod 8$			        &0 	&1	&2	&3	&4	&5	&6	&7	\\
\hline
$q\pmod4=1,~~ H_q $	      & 1+1	&  $ 1+1 $ 	& 1+1	& $ 1+1+ 1+1 $ 	& 2+2	& $  2+2 $	& 2+2	
& $ 1+1+ 1+1  $	\\

$q\pmod4=3, ~~ H_q $	& 1+1	& $ 1+1+1+1 $	& 2+2	& $ 2+2  $	& 2+2	& $ 1+1+1+1 $	& 1+1	&  $ 1+1 $	\\
\toprule
\end{tabular}
\label{tab:SUSYdeg}
\end{table}

 For $\mathcal{N}=1$ supersymmetric SYK model, for even $ N $ or odd $ N $ in the enlarged Hilbert space,
 $(-1)^F$ is always a conserved quantity $ [ H, (-1)^F ]=0 $.
 In the previous section, we classified the eigenstates of $ H_q= Q^2_q $ in terms of those of $ Q_q $.
 Due to $ \{ (-1)^F, Q_q \} =0 $, so  any eignestate of $ Q_q $ is NOT an eigen-state of the parity $ (-1)^F $,
 but a linear combination of the two with opposite parities \cite{quasi}.

 In this section, we will classify the eigenstate of $ H_q= Q^2_q $ in terms of parity $ (-1)^F $.
 Taking $ \psi_{+} $ as the ground state with a positive parity,
 Then $ H $ always has at least a doubly degenerate ground state with two opposite parities $ (\psi_{+}, Q \psi_{+} ) $.
 So in the following,  if the degeneracy at a given parity sector is $ d $.
 Then the total degeneracy $ d_t=2d $ is in-dependent of which basis one is using, so it should be the same as that
 achieved in the last section using $ Q_q $ basis.

Note that the anti-unitary operators $P$, $R$, $Z$ ( if exists ) and the fermion number parity $(-1)^F$  only depend on $N$.
So the commutation relations between these anti-unitary operators and the parity listed in  Eq.\eqref{eqSM:PRZF}
are independent of choice of Hamiltonian. The squared values of these anti-unitary operators Eq.\ref{squared} also hold
independent of choice of Hamiltonian.

\bigskip
{\bf D1. The systematic approach. }
\bigskip

Because $ q $ is fixed to be odd, so
in the following, we discuss even $ N $ and odd $ N $ respectively.

{\sl 1. $ N $ is even }

When $N$ is even, if $\psi_+$ is a even parity state,
then we need discuss the relations among all the states
generated from all the anti-unitary symmetry operators acting on $\psi_+$: $\psi_+, P\psi_+, R\psi_+, PR\psi_+ $ and also
their corresponding opposite parity partners  $Q_q\psi_+, PQ_q\psi_+, RQ_q\psi_+, PRQ_q\psi_+$.
From $ \Lambda=PR\propto(-1)^F$, one can see the relations
$\psi_+\propto PR\psi_+$, $P\psi_+\propto R\psi_+$ and also the relations among their opposite parity partners
$Q_q\psi_+\propto PRQ_q\psi_+$, $PQ_q\psi_+\propto RQ_q\psi_+$.
So one only need to clarify the relations among  the four states $\psi_+, P\psi_+$, and their corresponding opposite parity
partners $ Q_q\psi_+, PQ_q\psi_+ $.

%To tell if they are the same or not,
%we need knowledge on squared value of two anti-unitary operators $P^2$ and $(PQ_q)^2$,
%and $Q_q\psi_+$ always orthogonal to $\psi_+$ (because $Q_q$ always change parity).
%[Similar to $P^2$, if $PQ_q$ do NOT swap parity,
%then $(PQ_q)^2=+Q_q^2$ indicate $\psi_+\propto PQ_q\psi_+$,
%and $(PQ_q)^2=-Q_q^2$ indicate $\psi_+\not\propto PQ_q\psi_+$;
%if $PQ_q$ do swap parity, then $\psi_+\not\propto PQ_q\psi_+$.
%Similar rules will work for other antiuntary operator like $ZQ_q$.]
%We also need another important result of $\mathcal{N}=1$ SUSY SYK models,
%which ensured no SUSY broken at finite $N$.
%However, if one ignore this, then our counting of degeneracy will only work for none zero eigenenergy.

%The proof for ``$(PQ_q)^2=-Q_q^2$ leads to double degeneracy''.
%Consider $\psi_+$ state with eigenenergy $E$, so $Q_q^2|\psi_+\rangle=E|\psi_+\rangle$.
%If $PQ_q$ swaps parity then it is obvious lead to $\langle\psi_+|PQ_q\psi_+\rangle$.
%If $PQ_q$ preserves parity but still $[PQ_q,H]=0$, then
%\begin{align}
%    \langle\psi_+|PQ_q\psi_+\rangle
%	=(1/E)\langle PQ\psi_+|(PQ_q)^2\psi_+\rangle^*
%	=(1/E)\langle PQ\psi_+|(-Q_q^2)\psi_+\rangle^*
%	=-\langle \psi_+| PQ\psi_+\rangle
%\end{align}
%thus as long as $E>0$, $\langle\psi_+|PQ_q\psi_+\rangle$ must be zero,
%which indicate double degeneracy with $\psi_+\not\propto PQ_q\psi_+$.
%Alternatively, one can define a normalized $Q_q$ as $\hat{Q}_q=Q_q/\sqrt{E}$,
%then $(P\hat{Q}_q)^2=-1$ means double degeneracy.
%But this $\hat{Q}_q$ will depend on $E$ thus dependent on choose of $\psi_+$.

{\sl 2. $ N $ is odd }

When $ N $ is odd, if $\psi_+$ is a even parity state,
then we need discuss relation among all the states
generated from all symmetry operators acting on $\psi_+$,
$\psi_+$, $P\psi_+$, $R\psi_+$, $Z\psi_+$,
$PR\psi_+$, $ZR\psi_+$, $ZP\psi_+$, $ZPR\psi_+$ and their corresponding opposite parity partners
$Q_q\psi_+$, $PQ_q\psi_+$, $RQ_q\psi_+$, $ZQ_q\psi_+$,
$PRQ_q\psi_+$, $ZRQ_q\psi_+$, $ZPQ_q\psi_+$, $ZPRQ_q\psi_+$.
From $PR\propto(-1)^F$, one can see the relations
$\psi_+\propto PR\psi_+$, $P\psi_+\propto R\psi_+$,
$Z\psi_+\propto ZPR\psi_+$, $ZP\psi_+\propto ZR\psi_+$,
and also the relations among their opposite parity partners
$Q_q\psi_+\propto PRQ_q\psi_+$, $PQ_q\psi_+\propto RQ_q\psi_+$,
$ZQ_q\psi_+\propto ZPRQ_q\psi_+$, and $ZPQ_q\psi_+\propto ZRQ_q\psi_+$.
So one only need to clarify relations among the
8 states $\psi_+, P\psi_+, Z\psi_+, ZP\psi_+$ and their corresponding opposite parity partners
$Q_q\psi_+$, $PQ_q\psi_+$, $ZQ_q\psi_+$, and $ZPQ_q\psi_+$.
%In order to tell if they are the same or not,
%we need knowledge on squared value $P^2$ and $(ZQ_q)^2$ or $(PQ_q)^2$.

Now, it is important to observe that there is an extra conserved quantity
$ZPQ_q\propto\chi_\infty Q_q$  which also commutes with
the parity $[ZPQ_q,(-1)^F]=0$, so
one can construct the extra ``conserved quantity'' $(iZPQ_q)^\dagger=iZPQ_q$,
and $(iZPQ_q)^2=(i\chi_\infty Q_q)^2=\chi_\infty^2 Q_q^2=Q_q^2=H $.
Now the situation is similar to previous discussion in $[H,Q_q]$ basis where
one find an extra conserved quantity $ [ (-1)^F i \chi_{\infty}, Q_q]=0 $ in the $ Q_q $ basis.
Here one just shift $ \chi_{\infty} $ to $ Q_q $ to make the extra conserved quantity $ [ i\chi_\infty Q_q, (-1)^F ]=0 $
in the $ (-1)^F $ basis. In this basis, if choosing $\psi_{+,s}$ to satisfy $(-1)^F\psi_{+,s}= \psi_{+,s}$,
$iZPQ_q\psi_{+,s}=s\sqrt{E}\psi_{+,s}$  where $s=\pm$, then $H\psi_{+,s}=E\psi_{+,s}$.
The extra conserved quantities also leads to
$ZPQ_q\psi_{+,s} \sim \psi_{+,s}, ZQ_q\psi_{+,s} \sim P\psi_{+,s}, PQ_q\psi_{+,s} \sim Z\psi_{+,s}, ZP\psi_{+,s} \sim Q_q\psi_{+,s}  $, so
the 8 states reduce to 4 states $\psi_{+,+}$, $P\psi_{+,+}$, $Z\psi_{+,+}$, and $Q_q\psi_{+,+}$.
In the following, we will name the sign of eigenvalue of $iZPQ_q$ as $s$.

The relations among $P$, $R$, $Z$, $Q$ and $iZPQ_q$, can be easily worked out
\begin{alignat*}{2}
    &PiZPQ_qP^{-1}=(-1)^{\lfloor q/2\rfloor+1}(-1)^{(q+1)(N_c-1)}iZPQ_q,\quad
	&&RiZPQ_qR^{-1}=(-1)^{\lfloor q/2\rfloor+1}(-1)^{(q+1)N_c}iZPQ_q,\\
    &ZiZPQ_qZ^{-1}=(-1)^{\lfloor q/2\rfloor}(-1)^{(q+1)N_c}iZPQ_q,\quad
	&&Q_qiZPQ_qQ_q^{-1}=Q_qiZP=-iZPQ_q
\label{PRZQ}
\end{alignat*}
Thus for odd $ q $:
\begin{alignat*}{4}
    &PiZPQ_qP^{-1}=-iZPQ_q, \qquad
	&&ZiZPQ_qZ^{-1}=+iZPQ_q, \qquad
	    &&Q_qiZPQ_qQ_q^{-1}=-iZPQ_q, \qquad
		&&\text{if } q\pmod4=1\\
    &PiZPQ_qP^{-1}=+iZPQ_q, \qquad
	&&ZiZPQ_qZ^{-1}=-iZPQ_q, \qquad
	    &&Q_qiZPQ_qQ_q^{-1}=-iZPQ_q, \qquad
		&&\text{if } q\pmod4=3
%\label{PRZQ2}
\end{alignat*}
 where we dropped the $ R $ commutation relation related to Eq.\ref{PRZQ}, because, as demonstrated above,
we only need to tell the relations among $\psi_{+,+}$, $P\psi_{+,+}$, $Z\psi_{+,+}, Q_q\psi_{+,+}$.
So $P$, $Z$, and $Q_q$ are enough.

%\noindent\rule{\textwidth}{1pt}

Here, it is easy to see that the degeneracy of $H$ in extended scheme
is twice that of the degeneracy of $H$ in minima scheme,
due to two mutually commuting  parities.

In the following, we are ready to discuss the classification and level degeneracy.
All the results in the following sections D2-D5 are listed in Table \ref{tab:SUSYdeg}.

\bigskip
{\bf D2. Case $q\pmod4=1$ and even $ N $ }
\bigskip

When $N\pmod 8=0$, $N_c\pmod 4=0$, $P$ and $R$ preserve parity,
$\{P,Q_q\}=[R,Q_q]=0$, and $P^2=R^2=+1$.
%thus $Q_q$ belongs to Class BDI(chGOE).
If $\psi_+$ is an eigenstate of $H=Q_q^2$ with eigenenergy $E$ and even parity,
then $ P\psi_+ \propto \psi_+   $ is the same eigenstate
with eigenenergy $E$ and even parity,
$Q_q\psi_+\propto PQ_q\psi_+$ is another eigenstate
with eigenenergy $E$ and odd parity,
thus $H=Q_q^2$ has 1+1-fold degeneracy.

When $N\pmod 8=2$, $N_c\pmod 4=1$, $P$ and $R$ swap parity,
$[P,Q_q]=\{R,Q_q\}=0$, and $P^2=-R^2=+1$ and $(PQ_q)^2= Q_q^2$.
%thus $Q_q$ belongs to Class CI(chGOE).
If $\psi_+$ is an eigenstate of $H=Q_q^2$ with eigenenergy $E$ and even parity,
then $ PQ_q\psi_+ \propto \psi_+ $ is the same eigenstate
with eigenenergy $E$ and even parity,
$Q_q\psi_+\propto P\psi_+$ is another eigenstate
with eigenenergy $E$ and odd parity,
thus $H=Q_q^2$ has 1+1-fold degeneracy.

When $N\pmod 8=4$, $N_c\pmod 4=2$, $P$ and $R$ preserve parity,
$\{P,Q_q\}=[R,Q_q]=0$, and $P^2=R^2=-1$.
%thus $Q_q$ belongs to Class CII(chGSE).
If $\psi_+$ is an eigenstate of $H=Q_q^2$ with eigenenergy $E$ and even parity,
then $\psi_+$ and $P\psi_+$ are two orthogonal eigenstates
with eigenenergy $E$ and even parity,
$Q_q\psi_+$ and $PQ_q\psi_+$ are two orthogonal eigenstates
with eigenenergy $E$ and odd parity,
thus $H=Q_q^2$ has 2+2-fold degeneracy.

When $N\pmod 8=6$, $N_c\pmod 4=3$, $P$ and $R$ swap parity,
$[P,Q_q]=\{R,Q_q\}=0$, and $P^2=-R^2=-1$ and $(PQ_q)^2=-Q_q^2$.
% $Q_q$ belongs to Class A(GUE).
If $\psi_+$ is an eigenstate of $H=Q_q^2$ with eigenenergy $E$ and even parity,
then $\psi_+$ and $PQ_q\psi_+$ are two orthogonal eigenstates
with eigenenergy $E$ and even parity,
( This is because $ ( \psi_+, PQ_q\psi_+ )= ( PQ_q \psi_+, (PQ_q)^2 \psi_+ )^{*}/E= -( Q_q^2 \psi_+, PQ_q \psi_+)/E= -( \psi_+, PQ_q\psi_+ )=0 ) $, $Q_q\psi_+$ and $P\psi_+$ are two orthogonal eigenstates
with eigenenergy $E$ and odd parity,
thus $H=Q_q^2$ has 2+2-fold degeneracy.

\bigskip
{\bf D3. Case $q\pmod4=3$ and even $ N $ }
\bigskip

When $N\pmod 8=0$, $N_c\pmod 4=0$, $P$ and $R$ preserve parity,
$[P,Q_q]=\{R,Q_q\}=0$, and $P^2=R^2=+1$.
%thus $Q_q$ belongs to Class BDI(chGOE).
If $\psi_+$ is an eigenstate of $H=Q_q^2$ with eigenenergy $E$ and even parity,
then $ P\psi_+  \propto \psi_+ $ is the same eigenstate
with eigenenergy $E$ and even parity, $Q_q\psi_+\propto PQ_q\psi_+$ is an eigenstate
with eigenenergy $E$ and odd parity, thus $H=Q_q^2$ has 1+1-fold degeneracy.

When $N\pmod 8=2$, $N_c\pmod 4=1$, $P$ and $R$ swap parity,
$\{P,Q_q\}=[R,Q_q]=0$, and $P^2=-R^2=+1$ and $(PQ_q)^2=-Q_q^2$.
%thus $Q_q$ belongs to Class DIII(BdG).
If $\psi_+$ is an eigenstate of $H=Q_q^2$ with  eigenenergy $E$ and even parity,
then $\psi_+$ and $PQ_q\psi_+$ are two orthogonal  eigenstates
with eigenenergy $E$ and even parity,
$Q_q\psi_+$ and $P\psi_+$ are two orthogonal eigenstates
with eigenenergy $E$ and odd parity,
thus $H=Q_q^2$ has 2+2-fold degeneracy.

When $N\pmod 8=4$, $N_c\pmod 4=2$, $P$ and $R$ preserve parity,
$[P,Q_q]=\{R,Q_q\}=0$, and $P^2=R^2=-1$.
%thus $Q_q$ belongs to Class CII(chGSE).
If $\psi_+$ is an eigenstate of $H=Q_q^2$ with eigenenergy $E$ and even parity,
then $\psi_+$ and $P\psi_+$ are two orthogonal eigenstates
with eigenenergy $E$ and even parity,
$Q_q\psi_+$ and $PQ_q\psi_+$ are two orthogonal eigenstates
with eigenenergy $E$ and odd parity,
thus $H=Q_q^2$ has 2+2-fold degeneracy.

When $N\pmod 8=6$, $N_c\pmod 4=3$, $P$ and $R$ swap parity,
$\{P,Q_q\}=[R,Q_q]=0$, and $P^2=-R^2=-1$ and $(PQ_q)^2=Q_q^2$.
%thus $Q_q$ belongs to Class A(GUE).
If $\psi_+$ is an eigenstate of $H=Q_q^2$ with  eigenenergy $E$ and even parity,
then $ PQ_q\psi_+ \propto \psi_+ $ is the same eigenstate
with eigenenergy $E$ and even parity,
$Q_q\psi_+\propto P\psi_+$ is an eigenstate
with eigenenergy $E$ and odd parity,
thus $H=Q_q^2$ has 1+1-fold degeneracy.

\bigskip
{\bf D4. Case $q\pmod4=1$ and odd $ N $ }
\bigskip

Note that $P$ always maps $s\to-s$, $Z$ always maps $s\to +s$, $Q_q$ always maps $s\to -s$.
In this section, we will classify the energy levels in the parity basis of $ ( (-1)^F, i \chi_{\infty} Q_q ) $
which splits the Hilbert space into 4 sectors.

When $N\pmod 8=1$, $N_c\pmod 4=1$,
$P$ and $R$ swap parity, $Z$ preserve parity,
$[P,Q_q]=\{R,Q_q\}=\{Z,Q_q\}=0$,
and $P^2=-R^2=Z^2=+1$ and $(PQ_q)^2=+Q_q^2$.
If $\psi_{+,+}$ is an eigenstate of $H=Q_q^2$ with eigenenergy $E$, even parity and $s=+$,
then $Z\psi_{+,+}\propto \psi_{+,+}$
is an eigenstate with eigenenergy $E$, even parity and $s=+$,
$Q_q\psi_{+,+}\propto P\psi_{+,+}=\psi_{-,-}$
is an eigenstate with  eigenenergy $E$, odd parity and $s=-$,
thus $H=Q_q^2$ has $ 1_{+,+}+1_{-,-} $-fold degeneracy.

When $N\pmod 8=3$, $N_c\pmod 4=2$,
$P$ and $R$ preserve parity, $Z$ swaps parity,
$\{P,Q_q\}=[R,Q_q]=[Z,Q_q]=0$,
and $P^2=R^2=-Z^2=-1$.
If $\psi_{+,+}$ is an eigenstate of $H=Q_q^2$ with eigenenergy $E$,  even parity and $s=+$,
then $P\psi_{+,+}=\psi_{+,-}$ is an eigenstate of $H=Q_q^2$ with eigenenergy $E$, even parity and $s=-$,
$Z\psi_{+,+}=\psi_{-,+}$ is an eigenstate of $H=Q_q^2$ with eigenenergy $E$, odd parity and $s=+$,
$Q\psi_{+,+}=\psi_{-,-}$ is an eigenstate of $H=Q_q^2$ with  eigenenergy $E$, odd parity and $s=-$,
thus $H=Q_q^2$ has $ 1_{+,+}+1_{+,-}+ 1_{-,+}+1_{-,-} $-fold degeneracy.

When $N\pmod 8=5$, $N_c\pmod 4=3$,
$P$ and $R$ swap parity, $Z$ preserves parity,
$[P,Q_q]=\{R,Q_q\}=\{Z,Q_q\}=0$,
and $P^2=-R^2=Z^2=-1$ and $(PQ_q)^2=-Q_q^2$.
If $\psi_{+,+}$ is an eigenstate of $H=Q_q^2$ with eigenenergy $E$, even parity and $s=+$,
then
$\psi_{+,+}$ and $Z\psi_{+,+}=\tilde{\psi}_{+,+}$ are two orthogonal eigenstates
of $H=Q_q^2$ with eigenenergy $E$, even parity and $s=+$,
$P\psi_{+,+}=\psi_{-,-}$ and $Q\psi_{+,+}=\tilde{\psi}_{-,-}$ are two orthogonal eigenstates
of $H=Q_q^2$ with eigenenergy $E$, odd parity and $s=-$,
thus $H=Q_q^2$ has $ 2_{+,+}+2_{-,-} $-fold degeneracy.

When $N\pmod 8=7$, $N_c\pmod 4=0$,
$P$ and $R$ preserve parity, $Z$ swaps parity,
$\{P,Q_q\}=[R,Q_q]=[Z,Q_q]=0$,
and $P^2=R^2=-Z^2=+1$.
In this case,
if $\psi_{++}$ is an eigenstate of $H=Q_q^2$ with eigenenergy $E$, even parity and $s=+$,
then
$P\psi_{+,+}=\psi_{+,-}$ is an eigenstate of $H=Q_q^2$ with eigenenergy $E$, even parity and $s=-$,
$Z\psi_{+,+}=\psi_{-,+}$ is an eigenstate of $H=Q_q^2$ with eigenenergy $E$, odd parity and $s=+$,
$Q\psi_{+,+}=\psi_{-,-}$ is an eigenstate of $H=Q_q^2$ with eigenenergy $E$, odd parity and $s=-$,
thus $H=Q_q^2$ has $ 1_{+,+}+1_{+,-}+ 1_{-,+}+1_{-,-} $-fold degeneracy.

\bigskip
{\bf D5. $q\pmod4=3$ and odd $ N $ }
\bigskip

Note that $P$ always maps $s\to+s$, $Z$ always maps $s\to -s$, $Q_q$ always maps $s\to -s$.
In this section, we will classify the energy levels in the parity basis of $ ( (-1)^F, i \chi_{\infty} Q_q ) $
which splits the Hilbert space into 4 sectors.

When $N\pmod 8=1$, $N_c\pmod 4=1$,
$P$ and $R$ swap parity, $Z$ preserve parity,
$\{P,Q_q\}=[R,Q_q]=[Z,Q_q]=0$,
and $P^2=-R^2=Z^2=+1$.
If $\psi_{+,+}$ is an eigenstate of $H=Q_q^2$ with eigenenergy $E$, even parity and $s=+$,
then
$P\psi_{+,+}=\psi_{-,+}$ is an eigenstate of $H=Q_q^2$ with eigenenergy $E$, odd parity and $s=+$,
$Z\psi_{+,+}=\psi_{+,-}$ is an eigenstate of $H=Q_q^2$ with eigenenergy $E$, even parity and $s=-$,
$Q\psi_{+,+}=\psi_{-,-}$ is an eigenstate of $H=Q_q^2$ with eigenenergy $E$,  odd parity and $s=-$,
thus $H=Q_q^2$ has $ 1_{+,+}+1_{+,-}+ 1_{-,+}+1_{-,-} $-fold degeneracy.

When $N\pmod 8=3$, $N_c\pmod 4=2$,
$P$ and $R$ preserve parity, $Z$ swaps parity,
$[P,Q_q]=\{R,Q_q\}=\{Z,Q_q\}=0$,
and $P^2=R^2=-Z^2=-1$ and $(ZQ_q)^2=-Q_q^2$.
If $\psi_{+,+}$ is an eigenstate of $H=Q_q^2$ with eigenenergy $E$, even parity and $s=+$,
then
$\psi_{+,+}$ and $P\psi_{+,+}=\tilde{\psi}_{+,+}$ are two orthogonal eigenstates
of $H=Q_q^2$ with eigenenergy $E$, even parity and $s=+$,
$Z\psi_{+,+}=\psi_{-,-}$ and $Q\psi_{+,+}=\tilde{\psi}_{-,-}$ are two orthogonal eigenstates
of $H=Q_q^2$ with  eigenenergy $E$, odd parity and $s=-$,
thus $H=Q_q^2$ has  $  2_{+,+}+2_{-,-} $-fold degeneracy.

When $N\pmod 8=5$, $N_c\pmod 4=3$,
$P$ and $R$ swap parity, $Z$ preserves parity,
$\{P,Q_q\}=[R,Q_q]=[Z,Q_q]=0$,
and $P^2=-R^2=Z^2=-1$.
If $\psi_{+,+}$ is an eigenstate of $H=Q_q^2$ with eigenenergy $E$, even parity and $s=+$,
then
$P\psi_{+,+}=\psi_{-,+}$ is an eigenstate of $H=Q_q^2$ with eigenenergy $E$, odd parity and $s=+$,
$Z\psi_{+,+}=\psi_{+,-}$ is an eigenstate of $H=Q_q^2$ with eigenenergy $E$, even parity and $s=-$,
$Q\psi_{+,+}=\psi_{-,-}$ is an eigenstate of $H=Q_q^2$ with odd parity, $s=-$, and eigenenergy $E$,
thus $H=Q_q^2$ has $ 1_{+,+}+1_{+,-}+ 1_{-,+}+1_{-,-} $-fold degeneracy.

When $N\pmod 8=7$, $N_c\pmod 4=0$,
$P$ and $R$ preserve parity, $Z$ swaps parity,
$[P,Q_q]=\{R,Q_q\}=\{Z,Q_q\}=0$,
and $P^2=R^2=-Z^2=+1$ and $(ZQ_q)^2=+Q_q^2$.
If $\psi_{+,+}$ is an eigenstate of $H=Q_q^2$ with eigenenergy $E$, even parity and $s=+$,
then $P\psi_{+,+}\propto \psi_{+,+}$
is an eigenstate with eigenenergy $E$, even parity and $s=+$,
$Q_q\psi_{+,+}\propto Z\psi_{+,+}=\psi_{-,-}$
is an eigenstate with eigenenergy $E$, odd parity and $s=-$,
thus $H=Q_q^2$ has $ 1_{+,+}+1_{-,-} $-fold degeneracy.

\bigskip
{\bf D6. Comparison of Table \ref{tab:SYKq}  in the main text and Tables  \ref{tab:ext}, \ref{tab:SUSYdeg}. }
\bigskip

 When $q\pmod4=1 $,  at even $ N=0,2,4,6 $, $ d_H=2,2,4,4 $ in $ Q_q $ basis in Table \ref{tab:SYKq}
 which matches $ d_H=1+1, 1+1,2+2, 2+2  $ in the $ (-1)^{F} $ basis in Table \ref{tab:SUSYdeg}. Because $ \{ Q_q, (-1)^F \} =0 $,
 so the eigen-states in the $ Q_q $ basis has no definite parity.
 However, the degeneracy should be independent of the basis.
 At odd $ N=1,3,5, 7 $, $ d_H=2,4,4,4 $ in the extended $ ( Q_q, (-1)^{F} i\chi_\infty ) $ basis in Table \ref{tab:ext}
 matches $ d_H=1+1, 1+1+1+1,2+2, 1+1+1+1  $ in the $ ( (-1)^{F},  i\chi_\infty Q_q )  $ basis  Table \ref{tab:SUSYdeg}.
 Again, because $ \{ Q_q, (-1)^F \} =0 $, so the eigen-states in the former are different than those in the latter \cite{footnote1}.
 However, the total degeneracy should be independent of the basis.	

 When $q\pmod4=3 $,  at even $ N=0,2,4,6 $, $ d_H=2,4,4,2 $ in $ Q_q $ basis in Table \ref{tab:SYKq}
 matches $ d_H=1+1, 2+2,2+2, 1+1  $ in the $ (-1)^{F} $ basis in Table \ref{tab:SUSYdeg}. Because $ \{ Q_q, (-1)^F \} =0 $,
 so the eigen-states in the $ Q_q $ basis has no definite parity \cite{footnote1}.
% However, the degeneracy should be independent of the basis.
 At odd $ N=1,3,5, 7 $, $ d_H=4,4,4,2 $ in the extended $ ( Q_q, (-1)^{F} i\chi_\infty ) $ basis in Table \ref{tab:ext}
 matches $ d_H=1+1+1+1,2+2, 1+1+1+1, 1+1  $ in the extended $ ( (-1)^{F},  i\chi_\infty Q_q )  $ basis  Table \ref{tab:SUSYdeg}.
 Again, because $ \{ Q_q, (-1)^F \} =0 $, so the eigen-states in the former are different than those in the latter.
 However, the total degeneracy should be independent of the basis.

\section{E. RMT Classification of complex fermion SYK models}

The Hamiltonian for the SYK model with complex fermions is
\begin{align}
    H_c=\sum J_{i_1i_2\cdots i_{q/2};i_{q/2+1}\cdots i_q}
	c_{i_1}^\dagger c_{i_2}^\dagger \cdots c_{i_{q/2}}^\dagger
	c_{i_{q/2+1}} \cdots c_{i_q}
	-\mu F
\end{align}
where $q$ is an even number,
$J_{i_1i_2\cdots i_{q/2},i_{q/2+1}\cdots i_q}$
is complex Gaussian random variable antisymmetry
in $i_1i_2\cdots i_{q/2}$ and $i_{q/2+q}\cdots i_q$
and $J_{i_1i_2\cdots i_{q/2};i_{q/2+1}\cdots i_q}=
J_{i_{q/2+1}\cdots i_q;i_1i_2\cdots i_{q/2}}^*$, $\mu$ is the chemical potential for the fermion number operator
$F=\sum_{i=1}^{N_c} c_i^\dagger c_i$. $H_c$ has one conserved quantity $F$,
thus acquire a block diagonal structure labeled by $F=0,1,\cdots,N_c$.

The anti-unitary operators $P$, $R$ defined in Eq.\eqref{eq:CPR} satisfy:
\begin{align}
    Pc_iP^{-1}=(-1)^{N_c-1} c_i^\dagger,\qquad
    Pc_i^\dagger P^{-1}=(-1)^{N_c-1} c_i,\qquad
    Rc_iR^{-1}=(-1)^{N_c} c_i^\dagger,\qquad
    Rc_i^\dagger R^{-1}=(-1)^{N_c} c_i
\end{align}
thus $P$ and $R$ always map $F$ to $N_c-F$.

One can also work out their squared values $P^2=(-1)^{\lfloor N_c/2\rfloor}$
and $R^2=(-1)^{\lfloor 3N_c/2\rfloor}$ and
\begin{align}
    PH_cP^{-1}
	=&(-1)^{q(N_c-1)}\sum J_{i_1i_2\cdots i_{q/2};i_{q/2+1}\cdots i_q}^*
	c_{i_1} c_{i_2} \cdots c_{i_{q/2}}
	c_{i_{q/2+1}}^\dagger \cdots c_{i_q}^\dagger
	-\mu (N_c-F)\\
	=&(-1)^{q(N_c-1)}(-1)^{q/2}\sum J_{i_{q/2+1}\cdots i_q;i_1i_2\cdots i_{q/2}}
	c_{i_{q/2+1}}^\dagger \cdots c_{i_q}^\dagger
	c_{i_1} c_{i_2} \cdots c_{i_{q/2}}
	-\mu (N_c-F)
\end{align}
  where $ q $ is even.

So $PH_cP^{-1}=(-1)^{q/2}H_c$ only holds in the half-filling sector
$F=N_c/2$ (canonical ensemble) or $\mu=0$ (grand-canonical ensemble).
Similarly, $RH_cR^{-1}=(-1)^{q/2}H_c$ also only holds in the half-filling sector.
%No anti-unitary operator and chirality operator
%can commute with non-half-filled $F\neq N_c/2$.
%exist in the non-half-filled sector $F\neq N_c/2$.

The classification of complex fermion SYK models is determined by $q$ and $N_c$.
When $N_c$ is odd, no half-filling sector exists, thus all sectors belong to Class A(GUE).
When $N_c$ is even, all the non-half-filling sectors still belong to Class A(GUE).
But in the half-filling sector, $[P,H]=0$ if $q\pmod 4=0$ and the classification
is given by $P^2$  or $R^2$  (both $=T_+^2$); $\{P,H\}=0$ if $q\pmod 4=2$ and the classification
is given by $P^2$ or $R^2$ ( both $=T_-^2$).
The classification results are listed in Table. \ref{tab:cSYK}.

\begin{table}[!htb]
\caption{The RMT classification and energy level degeneracy of complex fermion SYK models at even $ q $.
The results for $q\pmod 4=0$ was achieved in \cite{You2017,hybrid}.
But the results for $q\pmod 4=2$ are new. }
\begin{tabular}{ c|cccc }
\toprule
$N_c\pmod 4$				&0	&1	&2	&3  \\ \hline
$P^2$ value	    			&$+$    &$+$	&$-$	&$-$    \\
$R^2$ value	    			&$+$    &$-$    &$-$    &$+$    \\ \hline
$q\pmod 4=0$    half-filling		&AI	&---    &AII	&---    \\
degeneracy in $F=N_c/2$	&1  &---&2  &---\\
  non-half-filling		&A	&A	&A	&A      \\
degeneracy in $F\neq N_c/2$	&1  &1	&1  &1	\\ \hline
$q\pmod 4=2$    half-filling		&D	&---	&C	&---    \\
degeneracy in $F=N_c/2$	&1  &---&1  &---\\
   non-half-filling		&A	&A	&A	&A      \\ %\hline
degeneracy in $F\neq N_c/2$	&1  &1	&1  &1	\\
\toprule
\end{tabular}
\label{tab:cSYK}
\end{table}

Now we determine the energy level degeneracy.
If $\psi_{F}$ is a eigenstate with eigenenergy $E$ in sector $F$,
then the degeneracy can be determined by clarifying the relations between
$\psi_{F}$ and $P\psi_{F}$,
Note that due to $(-1)^F=(-1)^{\lceil N_c/2\rceil}PR$,
$PR\psi_{F}\propto\psi_{F}$ and $R\psi_{F}\propto P\psi_{F}$.

{\sl (1) When $q\pmod 4=0$ } 

In this case, $[P,H_c]=0$.
If $\psi_{F}$ is an eigenstate with eigenenergy $E$
in non-half-filling sector $F$ (namely, $F\neq N_c/2$),
then $P\psi_{F}$ is an eigenstate with eigenenergy $E$
in another non-half-filling sector $N_c-F$.
Thus $H_c$ has no degeneracy in non-half-filling sector.

If $N_c\pmod 4=0$ and $\psi$ is an eigenstate with eigenenergy $E$
in half-filling sector $F=N_c/2$,
then $P\psi\propto \psi$ is the same eigenstate as $\psi$
in half-filling sector $F=N_c/2$.
Thus $H_c$ has no degeneracy in half-filling sector.
If $N_c\pmod 4=2$ and $\psi$ is an eigenstate with eigenenergy $E$
in half-filling sector $F=N_c/2$,
then $\psi$ and $P\psi$ are two orthogonal eigenstates
with the same eigenenergy $E$ in half-filling sector $F=N_c/2$.
Thus $H_c$ has double degeneracy in half-filling sector.

%To consider the total degeneracy in Hamiltonian $H_c$ (all sectors),
%for eigenenergy with $F\neq N_c/2$ is double degenerate,
%for eigenenergy with $F= N_c/2$ is double degenerate only if $N_c\pmod 4=2$
%otherwise not degenerate.

{\sl (2) When $q\pmod 4=2$ }

In this case, $\{P,H_c\}=0$.
If $\psi_{F}$ is an eigenstate with eigenenergy $E$
in non-half-filling sector $F$ (namely, $F\neq N_c/2$),
then $P\psi_{F}$ is an eigenstate with eigenenergy $-E$
in another non-half-filling sector $N_c-F$.
Thus $H_c$ has no degeneracy in non-half-filling sector.

If $N_c\pmod 4=0$ or $N_c\pmod 4=2$, $\psi_{F}$ is an eigenstate with eigenenergy $E$
in half-filling sector $F=N_c/2$,
then $P\psi_{F}$ is an eigenstate with eigenenergy $-E$
also in half-filling sector $F=N_c/2$.
%If $N_c\pmod 4=2$ and $\psi_{F,n}$ is an eigenstate with eigenenergy $E_n$
%in half-filling sector $F=N_c/2$,
%then $P\psi_{F,n}$ is an eigenstate with eigenenergy $-E_n$.
Thus $H_c$ has no degeneracy in half-filling sector, no matter $N_c\pmod 4=0$ or $2$,
%To consider the total degeneracy in Hamiltonian $H_c$ (all sectors),
but enjoys a spectral mirror symmetry.

The  pattern of energy levels of $H_c$
is listed in Table \ref{tab:cSYK}.
When comparing with Table \ref{tab:SYKq}, as expected, it is easy to see that
at half-filling case in Table \ref{tab:cSYK}, the two tables match
in both classification and degeneracy when setting $ N_c=2 N $ in both $q\pmod 4=0$ and $q\pmod 4=2$.
It is tempting to put the complex SYK with even $ q $ and the $ {\cal N}=2 $  SUSY SYK with odd $ q $
on the same periodic table \ref{tab:cSYK}.
It was known that the complex fermion SYK may be due to a charged ( Reissner-Nordstrom ) black hole which still has
no angular momentum \cite{Sachdev2015}.
So a periodic table of complex SYK and $ {\cal N}=2 $  SUSY SYK  may also lead to a classification of charged
quantum black holes. However, in sharp contrast
to the $ {\cal N}= 1 $  SUSY SYK, the $ {\cal N}=2 $ SUSY is not broken at any finite $ N $,
the Witten index $ W $ is non-vanishing, there is an extensive ground state degeneracy equal to $ W $ at the exact zero energy.
It is also not known if this extensive zero mode characterized by the Witten index can be identified as the
topological zero modes in the RMT.
However, as elucidated in this work, there is no zero modes in
the periodic table \ref{tab:SYKq} of SYK and $ {\cal N}=1 $  SUSY SYK.


\begin{thebibliography}{99}
\bibitem{Clifford}
The $d$-dimensional gamma matrices are be constructed from Pauli matrices ($d=2$ case)
$\gamma_1^{(2)}=\sigma_x$, $\gamma_2^{(2)}=\sigma_y$, $\gamma_3^{(2)}=\sigma_z$,
and then iteratively build $d+2$ matrices by taking tensor product
$\gamma_k^{(d+2)}=\sigma_x\otimes\gamma_k^{(d)}$, $k=1,\cdots,d+1$,
$\gamma_{d+2}^{(d+2)}=\sigma_y\otimes I_{2^{d/2}}$,
$\gamma_{d+3}^{(d+2)}=\sigma_z\otimes I_{2^{d/2}}$,
where $d$ is an even number.
The $N$ Majorana can be represented by $\chi_k=\gamma_k^{2\lfloor N/2\rfloor}$
with $k=1,2,\cdots,N$.
Note that in order to compare with conventional normalization
$\{\chi_i,\chi_j\}=\delta_{ij}$,
we also rescale $C\to 2^{-q/2}C$ when using $\{\chi_i,\chi_j\}=2\delta_{ij}$.
\bibitem{footnote1}
Note that in the minimum scheme used in the main text, one can not choose odd $i$ to be pure imaginary and even $i$ to be real when $N$ is odd,
because the number of real matrices equals the number of imaginary matrices plus one. However, in the extended scheme to be discussed
in the SM section C, one can use either way.
\bibitem{quasi} Here,  $\psi$ is the eigenstate of the supercharge $Q_q$, because $ \{(-1)^F, Q_q\}=0 $,
 so it does not have a given parity.
 The degeneracy will also be done with the parity eigen-states in the SM section C and D.
\end{thebibliography}
\end{document}